\documentclass[useAMS,usenatbib]{mn2e}
\usepackage{graphicx}
\usepackage{epstopdf}
\usepackage{color}

\newcommand{\rxte}{{\it RXTE}}
\newcommand{\rxteasm}{{\it RXTE}/ASM}
\newcommand{\rxtepca}{{\it RXTE}/PCA}
\newcommand{\rxtehexte}{{\it RXTE}/HEXTE}
\newcommand{\integral}{{\it INTEGRAL}}
\newcommand{\integralisgri}{{\it INTEGRAL}/ISGRI}
\newcommand{\integraljemx}{{\it INTEGRAL}/JEM-X}
\newcommand{\swift}{{\it Swift}}
\newcommand{\swiftbat}{{\it Swift}/BAT}
\newcommand{\swiftxrt}{{\it Swift}/XRT}

\newcommand{\isis}{{ISIS}}

\newcommand{\heasoft}{{\sc heasoft}}
\newcommand{\osa}{{\sc osa}}

\newcommand{\comptt}{{\sc comptt}}
\newcommand{\compst}{{\sc compst}}
\newcommand{\thcomp}{{\sc thcomp}}
\newcommand{\compps}{{\sc compps}}
\newcommand{\reflect}{{\sc reflect}}
\newcommand{\eqpair}{{\sc eqpair}}
\newcommand{\diskbb}{{\sc diskbb}}
\newcommand{\phabs}{{\sc phabs}}
\newcommand{\pcfabs}{{\sc pcfabs}}
\newcommand{\egauss}{{\sc egauss}}
\newcommand{\edge}{{\sc edge}}
\newcommand{\belm}{{\sc belm}}
\newcommand{\bremss}{{\sc bremss}}

\title[The spectro-timing analysis of major flares in Cyg X-3]{2006 May-July Major Radio Flare episodes in Cygnus X-3: spectro-timing analysis of the X-ray data}

\author[K. I. I. Koljonen et al.]
{K.~I.~I.~Koljonen$^{1}$\thanks{email: karri@kurp.hut.fi}, M.~L.~McCollough$^{2}$, D.~C.~Hannikainen$^{3}$, 
\and R.~Droulans$^{4,5}$
\\
$^{1}$Aalto University Mets\"ahovi Radio Observatory, Mets\"ahovintie 114, FIN-02540 Kylm\"al\"a, Finland \\
$^{2}$Smithsonian Astrophysical Observatory, 60 Garden Street, Cambridge, MA 02138-1516, USA \\
$^{3}$Department of Physics and Space Sciences, Florida Institute of Technology, 150 W. University Blvd., Melbourne, FL 32901, USA \\
$^{4}$Universit\'e de Toulouse; UPS-OMP; IRAP;  Toulouse, France\\
$^{5}$CNRS; IRAP; 9 Av. colonel Roche, BP 44346, F-31028 Toulouse cedex 4, France}

\begin{document}

\pagerange{\pageref{firstpage}--\pageref{lastpage}}
\pubyear{2012}

\maketitle

\label{firstpage}

\begin{abstract} 

We analyse in detail the X-ray data of the microquasar Cygnus X-3 obtained during major radio flaring episodes in 2006 with multiple observatories. The analysis consists of two parts: probing the fast ($\sim$ 1 minute) X-ray spectral evolution with Principal Component Analysis followed by subsequent spectral fits to the time-averaged spectra ($\sim$ 3~ks). Based on the analysis we find that the overall X-ray variability during major flaring episodes can be attributed to two principal components whose evolution based on spectral fits is best reproduced by a hybrid Comptonization component and a bremsstrahlung or saturated thermal Comptonization component. The variability of the thermal component is found to be linked to the change in the X-ray/radio spectral state. In addition, we find that the seed photons for the Comptonization originate in two seed photon populations that include the additional thermal emission and emission from the accretion disc. The Comptonization of the photons from the thermal component dominates, at least during the major radio flare episode in question, and the Comptonization of disc photons is intermittent and can be attributed to the phase interval 0.2--0.4. The most likely location for Comptonization is in the shocks in the jet. 

\end{abstract}


\section{Introduction} \label{introduction}

Cygnus X-3 (Cyg X-3) is a well-known X-ray binary (XRB), located at a distance of $\sim$ 8--10~kpc \citep{predehl1,predehl2,dickey} in the plane of the Galaxy. Its discovery dates back to 1966 \citep{giacconi} but the true nature of the system remains elusive despite extensive multiwavelength observations throughout the years. The system is thought to harbour a black hole due to its spectral resemblance to other black hole XRB systems, such as GRS 1915$+$105 and XTE J1550$-$564 (see e.g. \citealt{szostek3,hjalmarsdotter2}),
although a neutron star cannot be ruled out (e.g. \citealt{vilhu2}). Distinct 4.8-hour orbital modulation, typical of low-mass XRBs, is evident in the X-ray \citep{parsignault} and infrared \citep{mason} lightcurves, and yet infrared spectroscopic observations indicate that the mass-donating companion is a Wolf-Rayet (WR) star \citep{keerkwijk}, making it a high-mass XRB. Thus, it is quite possible that the accretion disk is enshrouded partially or wholly in the strong stellar wind of the WR companion (\citealt{fender}; \citealt{szostek2}; \citealt{vilhu2}). Unlike most other XRBs, Cyg X-3 is relatively bright in the radio virtually all of the time. It undergoes giant radio outbursts (up to 20~Jy, \citealt{waltman2}) with strong evidence of relativistic jet-like structures \citep{molnar,schalinski,mio}. Following the discovery of transient $\gamma$-ray emission from Cyg X-3 \citep{tavani,fermi}, the 
$\gamma$-ray emission process has been attributed to the interaction of the stellar wind with the relativistic jet \citep{zdziarski2,dubus}.

Modelling the X-ray data of X-ray binaries and microquasars often leads to a problem of degeneracy, i.e. distinct multiple models fit the observed data equally well (see e.g. \citealt{nowak} for a recent take on modelling the X-ray spectra of Cyg X-1). Therefore, even if an apparently good fit is obtained between the data and the model, it does not necessarily imply a match between theory and physical reality. In order to make sense out of this degeneracy we need to take other data dimensions besides spectral into account, namely timing and/or polarisation. While we cannot yet access the polarisation dimension, X-ray timing data is readily available, and several methods have been developed to combine spectral and timing analyses (e.g. \citealt{vaughan}). \citep{remillard} classified the accretion states of X-ray binaries based on concurrently examining their X-ray spectral energy distributions and power spectra. In this paper, we employ principal component analysis (PCA, not to be confused with the Proportional Counter Array onboard \rxte\/, always referred to here as \rxtepca), one of the standard tools of time series analysis (see e.g. \citealt{jolliffe} for a review) that has recently been introduced in the analysis of the X-ray data of XRBs by \citet{malzac} and references therein. Combined with the spectral resolving power of X-ray detectors, X-ray lightcurves -- instead of representing only one energy band -- can be transformed into a time series of X-ray spectra. Analysing these with PCA yield the principal components that are responsible for the spectral variation. This imposes a second requirement for the X-ray spectral fits, so that in addition to fitting the spectra acceptably, the resulting fits also have to satisfy the spectral evolution inferred from the PCA. This extra requirement reduces greatly, if not completely, the degeneracy of simply using the results from the spectral fits in determining the emission components of the system. Of course, the PCA does not tell us what the exact physical components are that correspond to the principal components, but these can be inferred by e.g. examining the variability spectrum. Ultimately, the physical model can be deduced through trial-and-error by fitting the X-ray spectra and searching for parameters that correlate with the principal components. \citet{malzac} showed for Cyg X-1 that the principal components suggested a variation in parameters within one spectral model (\eqpair\/, \citealt{eqpair}, specifically the compactness of soft seed photons and the compactness of the total power supplied to the plasma). Below we show that in the case of Cyg X-3 the principal components point to a difference in the normalisation of distinct spectral components.

In this paper we present a detailed analysis of the X-ray data from \rxte\/, \integral\/ and \swift\/ of Cyg X-3 taken during the major flaring episodes that occurred in 2006 May--June. In Section \ref{observations} we describe the observations used in the analysis, Section \ref{principal} introduces the PCA in the analysis of multi-energy X-ray lightcurves, and Section \ref{results} presents the results from the PCA as well as the results from modelling the X-ray spectra. The implications of these results are then discussed in Section \ref{discussion} and we present the conclusions of this paper in Section \ref{conclusions}.            

\section{Observations} \label{observations}

\begin{figure}
\begin{center}
\includegraphics[width=0.5\textwidth]{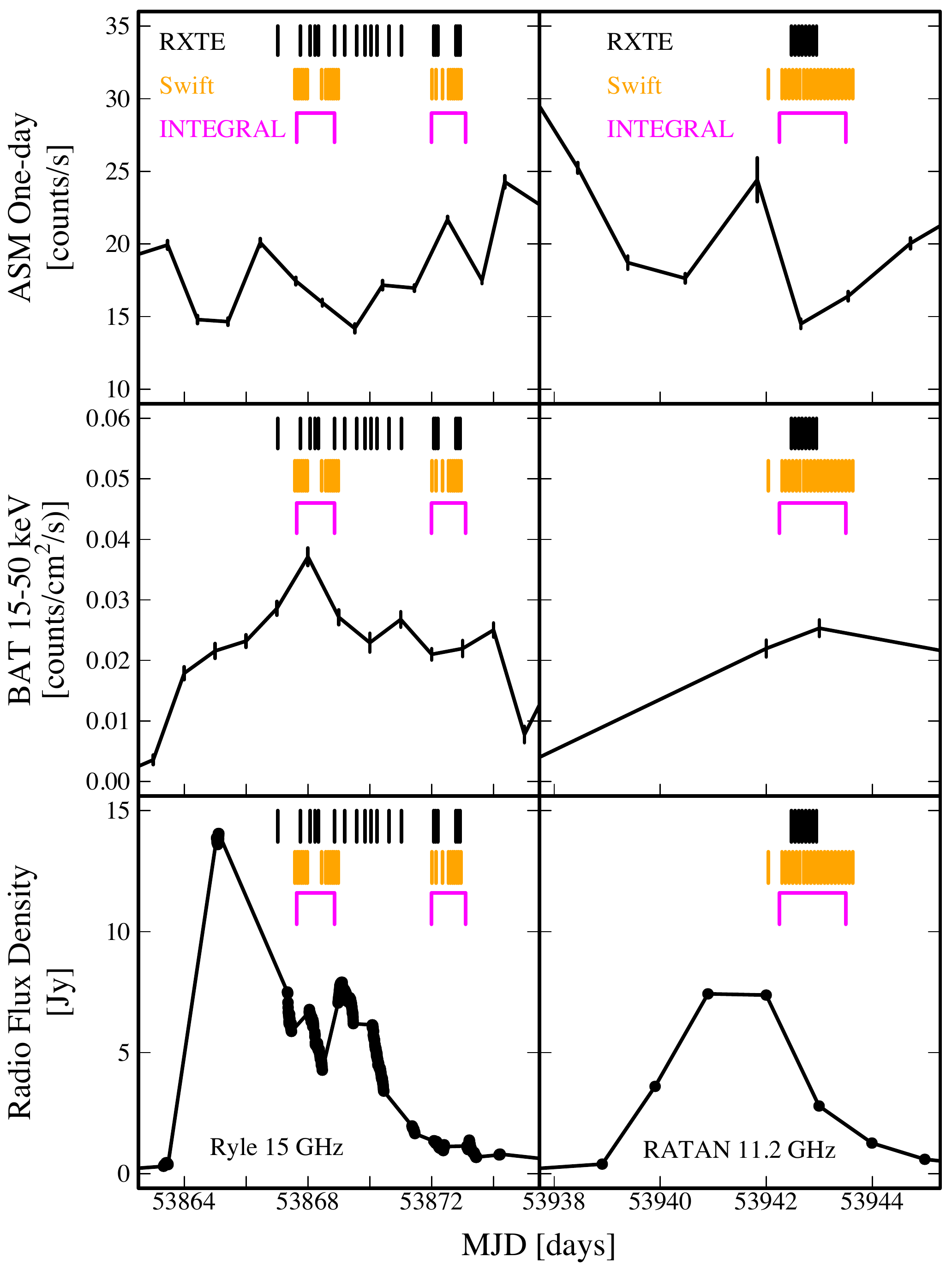}
\end{center}
\caption{The left-hand side panels from top to bottom show the one-day integrated \rxteasm\/, \swiftbat\/ and Ryle/AMI-LA 15~GHz lightcurves from the May 2006 major radio flare. The same is shown for the June 2006 major radio flare in the right-hand side panels, with instead the RATAN-600 radio lightcurve at 11.2~GHz. The top part of each panel shows the epochs of the \rxte\/ (black), \swift\/ (orange) and \integral\/ (magenta) data that are examined in this paper.} \label{obs}
\end{figure}

As mentioned in Section \ref{introduction}, the X-ray observations were obtained during the 2006 May--June major radio flaring episode, that consisted of two major radio flares with peak flux densities of 13.8~Jy (15~GHz) and 11.2~Jy (11.2~GHz) at MJD 53865 and MJD 53942, respectively (see Fig. \ref{obs}). For \rxte\/ (both \rxtepca\/ and \rxtehexte\/) and \swiftxrt\/ data we used the consequent observations taken right after a flare and for \integral\/ (\integraljemx\/ and \integralisgri\/) we used revolutions 437, 438 and 462 (see Tables \ref{swiftobs}, \ref{xobs} and \ref{intobs}). The radio data are from \citep{koljonen}.

We reduced all the data as described in the sections below. In addition to extracting X-ray spectra we constructed lightcurves from the data that will be used in the PCA. The orbital phase of individual segments of lightcurves were determined using a cubic ephemeris \citep{singh}.

\subsection{\swift}

The five \swiftxrt\/ observations used in this paper are tabulated in Table \ref{swiftobs}. Each \swift\/ observation covers approximately 12 hours and consists of several short pointings of $\sim$1~ks (which probe different phases of the Cyg X-3 orbit) for a total of up to 8~ks exposure times. The data were extracted using \heasoft\/ version 6.12 and reduced with the normal procedure using \textsc{xselect 2.4b}. The \swiftxrt\/ lightcurves for the PCA were extracted from 62 energy bands spanning an energy range from 1.1~keV to 8.0~keV, the first two bins having energy bin widths of 0.3~keV, the next three with energy bin widths of 0.2~keV, and the rest with energy bin widths of 0.1~keV. The background-subtracted 50~s binned lightcurve of the energy band 4.2--4.3~keV ranges from 0.6 to 3.7 counts/s with a mean error of 0.15 counts/s. Likewise, the lowest band used (1.1--1.4~keV) has mean counts of 0.5 counts/s with a mean error 0.09 counts/s, and the highest band used (7.9-8.0~keV) has mean counts of 0.12 counts/s with a mean error 0.04 counts/s. For the energy spectra we use the same energy range as for the lightcurves. In addition, we group the data eight bins to one and add 3\% systematic error for fitting purposes. 

\subsection{\rxte}

The 18 \rxte\/ observations used in this paper are tabulated in Table \ref{xobs}, with four longer observations divided into shorter segments (namely 91090-03-01-00, 91090-03-02-00, 91090-01-01-000 and 91090-01-01-00) resulting in approximately 3~ks long exposure times for \rxtepca\/ and 1~ks long exposure times for \rxtehexte\/. Each observation is individually reduced by the standard method as described in the \rxte\/ cook book using \heasoft\/ 6.12. The \rxtepca\/ lightcurves were extracted from energy bands 4--5, 5--6, 6--7, 7--9, 9--11, 11--13, 13--15 and 15--20~keV, whereas \rxtehexte\/ lightcurves were extracted from bands 20--30, 30--40, 40--50, 50--70, 70--100, 100--140 and 140--200~keV. During these observations only the \rxtehexte\/ Cluster B (Cl1) was available for lightcurve extraction and combined with the rocking of the instrument resulted in lightcurves that alternate between 16~s and 48~s intervals. Therefore we match the lightcurves from \rxtepca\/ to those from \rxtehexte\/ to obtain consistent lightcurves over the whole energy range of \rxte\/ for the PCA. For the \rxtepca\/, in the lowest band used (4--5~keV) the background-subtracted count rate ranges from 98.8 counts/s to 984.1 counts/s with a mean of 442.4 counts/s and a mean error of 5.1 counts/s, and in the highest band used (15--20~keV) the background-subtracted count rate ranges from 17.4 counts/s to 99.8 counts/s with a mean of 51.5 counts/s and a mean error of 1.8 counts/s. Likewise, for \rxtehexte\/ in the lowest band used (20--30~keV) the background-subtracted count rate ranges from 0.9 counts/s to 17.5 counts/s with a mean of 8.3 counts/s and a mean error of 1.2 counts/s, and in the highest band used (140--200~keV) the background-subtracted count rate ranges from $-$3.8 counts/s to 17.3 counts/s with a mean of 0.8 counts/s and a mean error of 1.0 counts/s. For the energy spectra we use the standard 128-channel \rxtepca\/ and 64-channel \rxtehexte\/ data with the same energy range as for the lightcurves. For fitting purposes we add 0.5\% systematic error to the \rxtepca\/ data. In addition, we group \rxtehexte\/ data to a minimum of six sigma significance.

\subsection{\integral}  

The three \integral\/ observations are summarised in Table \ref{intobs}. All three observations are approximately 100~ks long, comprising six orbital periods with small gaps resulting in the exposure times shown in Table \ref{intobs}. The observations were analysed using the \osa\/ version 9.0. The lightcurves for \integraljemx\/ and \integralisgri\/ were extracted from the same energy bands as for \rxte\/ with time bins of 100~s. The lowest band (4--5~keV) of \integraljemx\/ has a background-subtracted count rate that ranges from $-$2.5 counts/s to 29.0 counts/s with a mean of 8.4 counts/s and a mean error of 1.9 counts/s. Similarly, the background-subtracted count rate for highest band (15--20~keV) ranges from $-$9.3 counts/s to 18.7 counts/s with a mean of 1.8 counts/s and a mean error of 1.8 counts/s. Likewise, the lowest band (20--30~keV) of \integralisgri\/ has a background-subtracted count rate that ranges from $-$4.1 counts/s to 40.5 counts/s with a mean of 15.3 counts/s and a mean error of 3.5 counts/s. The highest band (140--200~keV) of \integralisgri\/ has a background-subtracted count rate that ranges from $-$6.8 counts/s to 5.9 counts/s with a mean of 0.1 counts/s and a mean error of 1.8 counts/s. For fitting the X-ray spectra we use the standard 64-channel \integraljemx\/ and 16-channel \integralisgri\/ data with a similar energy range as for \rxte\/. We added 3\% and 5\% systematic error to \integraljemx\/ and \integralisgri\/ respectively.

\begin{table*}
\caption{Log of \swift\/ observations} \label{swiftobs}
\begin{center}
\begin{tabular}{llccccc}
\hline\hline
N$^{\underline{o}}$ & ObsID & Date & MJD & N$^{\underline{o}}$ & Phase & XRT \\
& & & Interval & Pointings & Intervals & Exp. \\
& & [yy/mm/dd] & [d] & & [ks] \\
\hline
1 & 00053560010 & 06/05/12 & 53867.579--53867.990 & 7 & 0.18--0.24/0.51--0.57/0.85--0.91 & 5.64 \\
2 & 00053560011 & 06/05/13 & 53868.443--53868.992 & 8 & 0.17--0.25/0.51--0.59/0.85--0.92 & 7.37 \\
3 & 00053560012 & 06/05/17 & 53872.008--53872.949 & 10 & 0.00--0.07/0.34--0.41/0.67--0.74 & 7.86 \\
4 & 00053560013 & 06/07/26 & 53942.034--53942.975 & 12 & 0.00--0.02/0.06--0.09/0.34--0.41/0.67--0.74 & 3.81 \\
5 & 00053560014 & 06/07/27 & 53943.027--53943.636 & 10 & 0.00--0.05/0.34--0.39/0.67--0.72 & 5.91 \\
\hline
\end{tabular}
\end{center}
\end{table*}
     
\begin{table*}
\caption{Log of \rxte\/ observations} \label{xobs}
\begin{center}
\begin{tabular}{llccccccc}
\hline\hline
N$^{\underline{o}}$ & ObsID & Date & MJD & Phase & PCA & HEXTE & X-ray/ & Radio \\
& & & Interval & Interval & Exp. & Exp. & Radio & Flux \\
& & [yy/mm/dd] & [d] & & [ks] & [ks] & State & [Jy] \\
\hline
1 & 91090-02-01-00 & 06/05/12 & 53867.006--53867.040 & 0.980--1.146 & 2.8 & 1.0 & FHXR & 7.498 (Ry) \\
2 & 91090-02-01-06 & 06/05/12 & 53867.725--53867.777 & 0.580--0.766 & 3.2 & 1.1 & FIM & 6.348 (Ry) \\  
3 & 91090-02-01-01 & 06/05/13 & 53868.052--53868.093 & 0.219--0.407 & 3.2 & 1.1 & FHXR & 6.558 (Ry) \\
4 & 91090-02-01-07 & 06/05/13 & 53868.181--53868.271 & 0.875--1.312 & 5.1 & 1.5 & FHXR & 5.519 (Ry) \\
5 & 91090-02-01-02 & 06/05/13 & 53868.315--53868.354 & 0.531--0.719 & 3.2 & 0.9 & FHXR & 5.160 (Ry) \\
6 & 91090-02-01-05 & 06/05/13 & 53868.836--53868.894 & 0.154--0.342 & 3.2 & 1.1 & FHXR & 7.174 (Ry) \\ 
7 & 91090-02-01-03 & 06/05/13 & 53869.166--53869.214 & 0.798--0.982 & 3.2 & 1.0 & FHXR & 6.678 (Ry) \\
8 & 91090-02-01-08 & 06/05/14 & 53869.558--53869.599 & 0.761--0.898 & 2.3 & 0.8 & FIM & 6.316 (Ry) \\
9 & 91090-02-01-04 & 06/05/14 & 53869.820--53869.871 & 0.073--0.261 & 3.2 & 1.1 & FIM--FHXR & 5.745 (Ry) \\ 
10 & 91090-02-01-10 & 06/05/15 & 53870.017--53870.057 & 0.056--0.245 & 3.2 & 1.1 & FIM & 5.831 (Ry) \\
11 & 91090-02-01-12 & 06/05/15 & 53870.214--53870.254 & 0.041--0.229 & 3.2 & 0.9 & FIM & 4.829 (Ry) \\
12 & 91090-02-01-11 & 06/05/15 & 53870.603--53870.649 & 0.007--0.160 & 2.6 & 0.9 & FIM & 3.643 (Ry) \\
13 & 91090-02-01-09 & 06/05/15 & 53870.998--53871.039 & 0.975--1.164 & 3.3 & 1.1 & FIM & 1.869 (Ry) \\
14 & 91090-03-01-00(1) & 06/05/17 & 53872.051--53872.084 & 0.243--0.406 & 2.8 & 0.9 & FIM & 1.353 (Ry) \\
15 & 91090-03-01-00(2) & 06/05/17 & 53872.112--53872.149 & 0.545--0.735 & 3.3 & 1.0 & FIM & 1.275 (Ry) \\
16 & 91090-03-01-00(3) & 06/05/17 & 53872.177--53872.215 & 0.874--1.063 & 3.3 & 0.9 & FIM & 1.208 (Ry) \\
17 & 91090-03-02-00(1) & 06/05/17 & 53872.767--53872.804 & 0.825--1.015 & 3.3 & 1.1 & FIM & 1.120 (Ry) \\
18 & 91090-03-02-00(2) & 06/05/17 & 53872.832--53872.870 & 0.153--0.342 & 3.3 & 1.1 & FIM & 1.120 (Ry) \\
19 & 91090-03-02-00(3) & 06/05/17 & 53872.898--53872.935 & 0.481--0.670 & 3.3 & 1.1 & FIM & 1.120 (Ry) \\
20 & 91090-01-01-000(1) & 06/07/26 & 53942.459--53942.499 & 0.827--1.028 & 3.5 & 1.0 & FHXR & 2.797 (RA) \\
21 & 91090-01-01-000(2) & 06/07/26 & 53942.525--53942.565 & 0.155--0.356 & 3.5 & 1.0 & FHXR & 2.797 (RA) \\
22 & 91090-01-01-000(3) & 06/07/26 & 53942.590--53942.630 & 0.483--0.684 & 3.5 & 1.0 & FHXR & 2.797 (RA) \\
23 & 91090-01-01-000(4) & 06/07/26 & 53942.656--53942.696 & 0.811--1.012 & 3.5 & 1.0 & FHXR & 2.797 (RA) \\
24 & 91090-01-01-00(1) & 06/07/26 & 53942.729--53942.761 & 0.178--0.340 & 2.8 & 0.9 & FHXR & 2.797 (RA) \\
25 & 91090-01-01-00(2) & 06/07/26 & 53942.799--53942.827 & 0.528--0.668 & 2.4 & 0.8 & FHXR & 2.797 (RA) \\
26 & 91090-01-01-00(3) & 06/07/26 & 53942.868--53942.892 & 0.872--0.996 & 2.1 & 0.7 & FHXR & 2.797 (RA) \\
27 & 91090-01-01-01 & 06/07/26 & 53942.937--53942.959 & 0.222--0.327 & 1.8 & 0.6 & FHXR & 2.797 (RA) \\
\hline
\end{tabular}
\end{center}
\end{table*}

\begin{table*}
\caption{Log of \integral\/ observations} \label{intobs}
\begin{center}
\begin{tabular}{llcccc}
\hline\hline
N$^{\underline{o}}$ & Rev. & Date & MJD & JMX1 & ISGRI \\
& & & Interval & Exp. & Exp. \\
& & [yy/mm/dd] & [d] & [ks] & [ks] \\
\hline
1 & 437 & 06/05/12 & 53867.63--53868.90 & 75.4 & 60.9 \\
2 & 438 & 06/05/17 & 53872.00--51873.10 & 67.4 & 55.0 \\
3 & 462 & 06/07/26 & 53942.25--53943.50 & 52.9 & 64.2 \\
\hline
\end{tabular}
\end{center}
\end{table*}

\section{Principal Component Analysis} \label{principal}

The main use of PCA is to find the smallest number of components that are sufficient in describing the data while losing the least amount of information in the process.
Basically, PCA finds patterns in the data in a way that highlights the differences and similarities in the data set. In many-dimensional cases, where a graphical representation is not convenient, it is a particularly powerful analysis tool. By identifying the ``new coordinates" of the data set (i.e. the principal components) where the data points primarily cluster and ignoring the small scatter of the points around these coordinates, the dimensionality of the data set is greatly reduced, defined only by these new coordinates. In order to apply PCA to our X-ray spectra we closely followed the work of \citet{malzac} and the reader is highly recommended to read that paper and references therein for further details. Here, we briefly summarise the main points of the analysis.

We start with a number of spectra $p$ measured at times $t_{1}$, $t_{2}$, \ldots$t_{p}$ binned into $n$ bins corresponding to energies $E_{1}$, $E_{2}$, \ldots$E_{n}$ (see Section 2). The spectra are then viewed as $p \times n$ matrix with coefficients $F(t_{p},E_{n})$, i.e. the fluxes in each energy band at times $p$. For the PCA to work, one needs to subtract a mean flux $\bar{F}(E_{n})$ from the coefficients for each energy channel. One can compute the $n \times n$ covariance matrix out of the data matrix, which states the variances between each $n$-dimensions. The PCA proceeds to calculate the eigenvectors, $c_{n}$, and eigenvalues of the covariance matrix. These eigenvectors form the new coordinates of the data and the accompanying eigenvalue states the proportion of variance of a particular eigenvector, i.e. the highest eigenvalue and accompanying eigenvector is the first principal component of the data set etc. One can then form a linear decomposition $C_{nk}=(c_{n1}c_{n2}\ldots c_{nk})$ of the data set using these eigenvectors, ordered by the proportion of variance. The components producing only a small fraction of the overall variance can then be dropped, reducing the dimensionality, i.e. choosing a small number of eigenvectors for the linear composition (see more about choosing the number of components in Section \ref{lev}). The data set defined by these new coordinates can be obtained by

\begin{equation}
F_{kp} = C_{kn} \times \bar{F}_{np} = \sum_{i=1}^{n} C_{ki}\bar{F}_{ip} 
\end{equation}

where the chosen $k$ coordinates lie in the direction of the calculated eigenvectors $c_{k}$. Then one can work backwards producing the original data set without the less significant components, 

\begin{equation} \label{eqpca}
F_{np}^{PCA} = (C_{nk} \times F_{kp}) + \bar{F}(E_{n}) = \bar{F}(E_{n}) + \sum_{i=1}^{k} C_{ni}F_{ip}
\end{equation}

which most likely are just systematic errors and noise, leaving only the intrinsic variability of the data set. Just as easily and more importantly, one can concentrate on just one principal component, so as to isolate the effect caused by this component to the overall variability. In the following section we will exploit PCA in different ways using both spectral and timing information. In the spectral domain this includes producing average, minimum and maximum energy spectra out of individual principal components (Section \ref{exspec}), their fractions of variance as a function energy (Section \ref{varspec}), and their correlation with the measured flux. In the timing domain we can follow the individual principal components by tracking the evolution of $F_{kp}$ with time $t_{p}$, for each $k=1,2,3\ldots$ Similarly, we can test the correlation of $F_{kp}$ with the measured flux with time $t_{p}$ for each energy band $E_{n}$. 

\subsection{Log-Eigenvalue Diagram (LEV)} \label{lev}

The usual tool for deciding how many principal components one should retain is called the scree graph or log-eigenvalue (LEV) diagram (see e.g. \citealt{jolliffe} for a review). In the scree graph, the eigenvalues $l_{n}$ are plotted against $n$ and the critical point is where the eigenvalues below it are steep and those after it are flat, thus forming an ``elbow" in the graph. An alternative method -- and the one used in this paper -- is the LEV which plots the logarithm of $l_{n}$ against $n$. If the ``noise" components are decaying in a geometric progression, the corresponding eigenvalues will appear as a straight line in the LEV diagram. Also, instead of plotting the eigenvalues $l_{n}$, we use the proportion of variance, which is defined as $t_{k}= l_{k} / \sum_{i=1}^{n} l_{i}$, with $k=1,2,\ldots,n$.

\subsection{Extremal energy spectra} \label{exspec}

The contribution of an individual principal component to the energy spectrum can be determined by looking at the $F_{np}^{PCA}$ from Eq. \ref{eqpca} as a function of energy $E_{n}$ for each $k=1,2,3\ldots$ The minimum and maximum effect to the energy spectra is calculated by integrating $F_{np}^{PCA}$ over the energy range for each $p$ (and $k=1,2,3\ldots$), and then selecting the minimum and maximum value which then represent the minimum and maximum effect to the shape and normalisation of the energy spectrum.  

\subsection{Variance spectrum} \label{varspec}

The variance spectrum is a graph that shows the measured variance as a function of energy $E_{n}$. It can be plotted for all the principal components, i.e. showing the overall variance across the energy range of the data, and it is obtained as the diagonal components of the covariance matrix $C_{nn}$. On the other hand the variance spectrum can be calculated for the individual principal component as the diagonal of the covariance matrix of $F_{np}^{PCA}$ from Eq. \ref{eqpca}, for each $k=1,2,3\ldots$ One can then also calculate the contribution of each principal component to the total variance as a function of energy by dividing the individual variance spectra by the total variance spectrum.   

\section{Results} \label{results}

We divide this section into two parts, first concentrating on the PCA results from \swift\/, \integral\/ and \rxte, and then on the fitting of the \rxte\/ spectra.

\subsection{PCA results} \label{pcaresults}  

\begin{figure*}
\begin{center}
\includegraphics[width=1.0\textwidth]{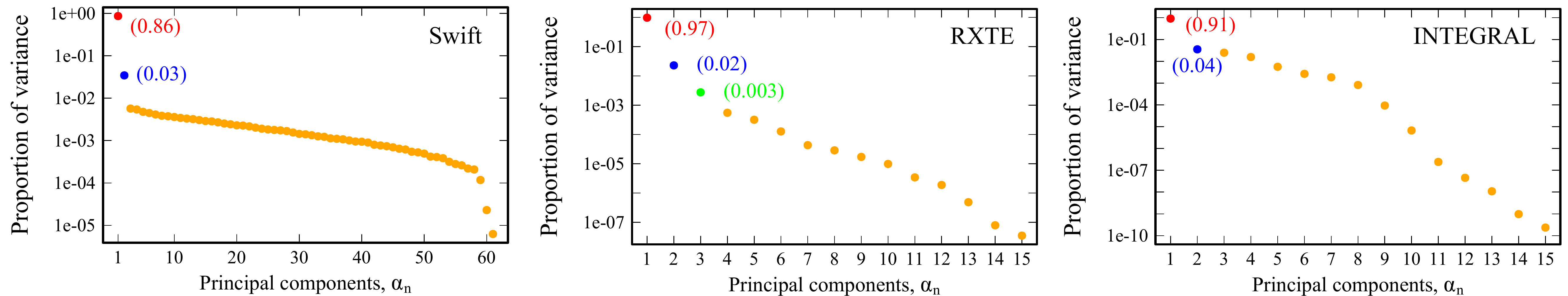}
\end{center}
\caption{The LEV diagrams of the principal components from all X-ray observatories (left: \swift\/, center: \rxte\/, right: \integral\/). The panels show the proportion of variance attributed to each principal component coloured corresponding to Fig. \ref{sigmaplot}. Also, the individual proportion of variance is labelled for the first two or three principal components that are presented in Figs. \ref{sigmaplot} and \ref{pca}.} \label{screeplot}
\end{figure*}

\begin{figure*}
\begin{center}
\includegraphics[width=1.0\textwidth]{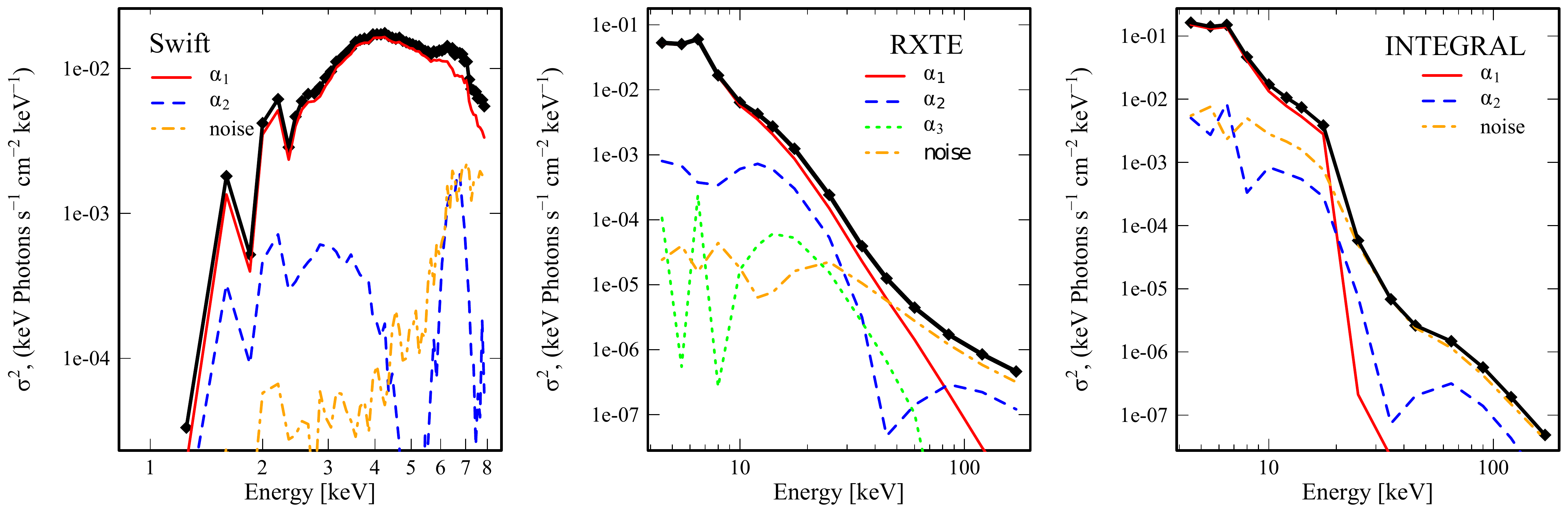}
\end{center}
\caption{The variance spectra of all observations (black solid line with data points) from all observatories (left: \swift\/, center: \rxte\/, right: \integral\/). The lower curves in the figure show the contributions of the principal components $\alpha_{1}$ (red, solid), $\alpha_{2}$ (blue, dash), $\alpha_{3}$ (green, dot) and the remainder of the components totalling to noise and systematic errors (orange, dot-dash).} \label{sigmaplot}
\end{figure*}

\begin{figure*}
\begin{center}
\includegraphics[width=1.0\textwidth]{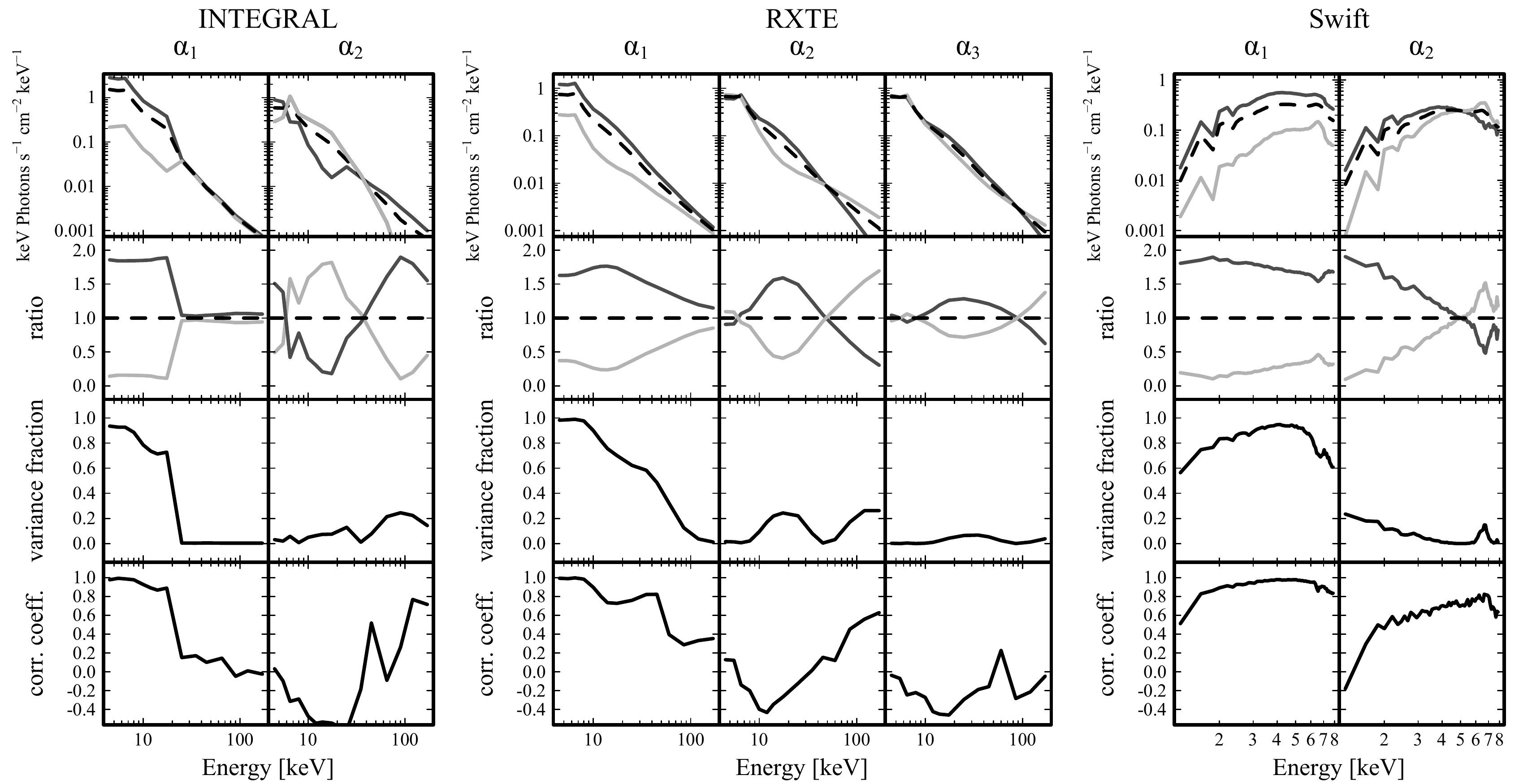}
\end{center}
\caption{The first two or three principal components ($\alpha_{1}$, $\alpha_{2}$ and $\alpha_{3}$) for data from all X-ray observatories. The upper panels show the average (dash), minimum and maximum (thick lines) spectra of the principal components. The upper middle panels show the ratio of minimum and maximum spectra to the average one. The lower middle panels shows the fraction of variance that the principal components accounts for as a function of energy and the bottom panels show the correlation coefficients of the principal components with flux in each energy band.} \label{pca}
\end{figure*}

The result of the PCA analysis is depicted in Figs. \ref{screeplot}, \ref{sigmaplot} and \ref{pca} separately for each X-ray observatory. Fig. \ref{screeplot} shows the fractions of variance attributed to the principal components. Fig. \ref{sigmaplot} shows the variance spectra for the three (\rxte\/) and the two (\swift/\integral) most significant principal components, as well as the sum of the remaining components labelled as noise. Fig. \ref{pca} shows what is the effect of the two or three most significant principal components on the Cyg X-3 spectra and in what energy bands the variance occurs, as well as how it is correlated to the flux in each energy band. Overall, the first principal component in all observatories contributes the most to the overall variance (86\%, 97\% and 91\% for \swiftxrt\/, \rxte\/ and \integral\/ respectively, see Fig. \ref{screeplot}), corresponds to the change in normalisation in the X-ray spectra, and is positively correlated with the measured flux and thus with the orbital modulation (Fig. \ref{pca}, bottom row). Since this component is modulated along with the X-ray fluxes, it means that the emission is a local one and, thus, not connected to the radio flux, which does not show any distinct modulation. In addition to the first principal component, there are also other, less significant, principal components that contribute in a small way to the overall variability.

In the following we will review the PCA results of each observatory independently.

\subsubsection{\swift} \label{swift}

For \swift\/, the first two principal components add up to nearly 90\% of the overall variance with most of the variance concentrated in the first principal component (86\%). Based on the LEV diagram (Fig. \ref{screeplot}), there is a sharp elbow after the two principal components indicating the start of the noise. Noticeable features in the variance spectrum (Fig. \ref{sigmaplot}) are dips in energy bins 1.8--1.9 and 2.3--2.4~keV and an increase of variability in the iron line region. The dips corresponds roughly to the location of some of the strongest emission lines in the X-ray spectra (H-like silicon at $\sim$ 2.0~keV and H-like sulphur at $\sim$ 2.5~keV) and could be interpreted as a reduction in variability indicating line production further out from the compact object in the photoionised stellar wind. However, the dips could also be due to a low count rate and wide energy bins. The decrease in variability in the second principal component around $\sim$5~keV as visible in the variance spectrum is most likely of instrumental origin since there are no known emission lines in that region. To confirm this, we performed a similar PCA analysis on a set of simulated lightcurves based on the best-fitting model as we did on the original data (see Section \ref{simulations}). Both principal components show a change in variability in the iron line region, so that a decrease in variability corresponds to times of maximum spectra and the increase to times of minimum spectra (Fig. \ref{pca}, upper and upper middle rows). Interestingly, the first principal component show just a peak around 6.5~keV corresponding to mostly cold iron line, while a sharper peak at 6.7~keV corresponding to ionised helium-like iron line in addition to 6.5~keV peak is present in the second principal component. This implies that there are two regions where the iron line forms, where the other (corresponding to the second principal component) is more ionised and thus hotter. Both principal components are correlated with the X-ray fluxes (Fig. \ref{pca}, bottom row), although the second principal component slightly less but the correlation is increasing towards the iron line region. Compared to the principal components of \rxte\/ and \integral\/, the variance spectrum of \swift\/ appears to not to include a third principal component. The analysis of the simulated data in Section \ref{simulations} indicates that the spectral components that were found to correspond to the second and third principal components seen in the \rxte\/ and \integral\/ data are clumped together in \swift\/ data, most likely an effect due to the small size of the dataset.  

\subsubsection{\rxte} 

For \rxte\/, almost all the variance (97\%) is due to the first principal component. Compared to the principal components from the \swift\/ analysis, it is more difficult to establish where the noise components commence based on the LEV diagram (Fig. \ref{screeplot}). However, some levelling off of the components is visible after the third principal component, and the fourth principal component dips below 0.1\%, a very small proportion of the variance and thus most likely contributing to the noise. The amount of \rxte\/ data compared to the other observatories is manifold, and thus we can expect the noise level to be lower. We opted to include the third principal component, albeit contributing only 0.3\% to the overall variance, since its shape in the variance spectrum (Fig. \ref{sigmaplot}) resembles a reflection spectrum. 

The second principal component contributes 2.3\% of the overall variance. The variance spectrum of this component exhibits a cut-off at $\sim$10~keV. Despite not contributing much to the variance overall -- 20\% at most around 20~keV (Fig. \ref{pca}, lower middle row) -- the fact that it affects the energy spectra most in the region 10--40~keV (Fig. \ref{pca}, upper and upper middle row) explains why it is required to fit the X-ray spectra (see Section \ref{model}). Interestingly, this component is mildly anti-correlated with the X-ray fluxes. Fig. \ref{pca} shows that the third component impacts the energy spectra and the variance minimally, and it is not correlated with the X-ray fluxes.

\begin{figure}
\begin{center}
\includegraphics[width=0.5\textwidth]{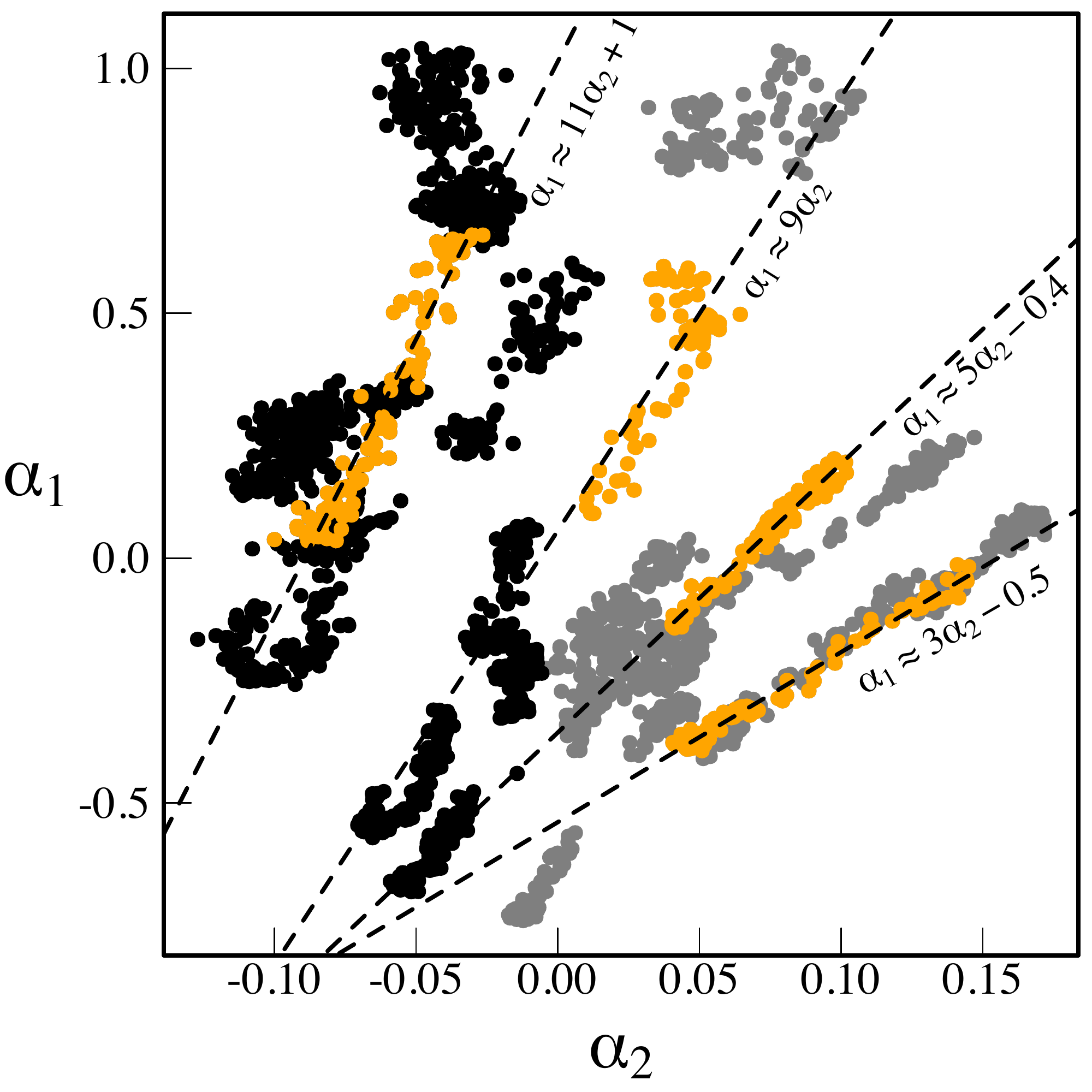}
\end{center}
\caption{The first principal component ($\alpha_{1}$) plotted as a function of the second principal component ($\alpha_{2}$) as calculated from the \rxte\/ data. Black and grey points refer to FHXR and FIM X-ray/radio state respectively as marked in Table \ref{xobs}. Orange points refer to pointings 7, 8, 10 and 17 from left to right with a simple linear regression fitted for each pointing.} \label{compcomp}
\end{figure}

\begin{figure}
\begin{center}
\includegraphics[width=0.35\textwidth]{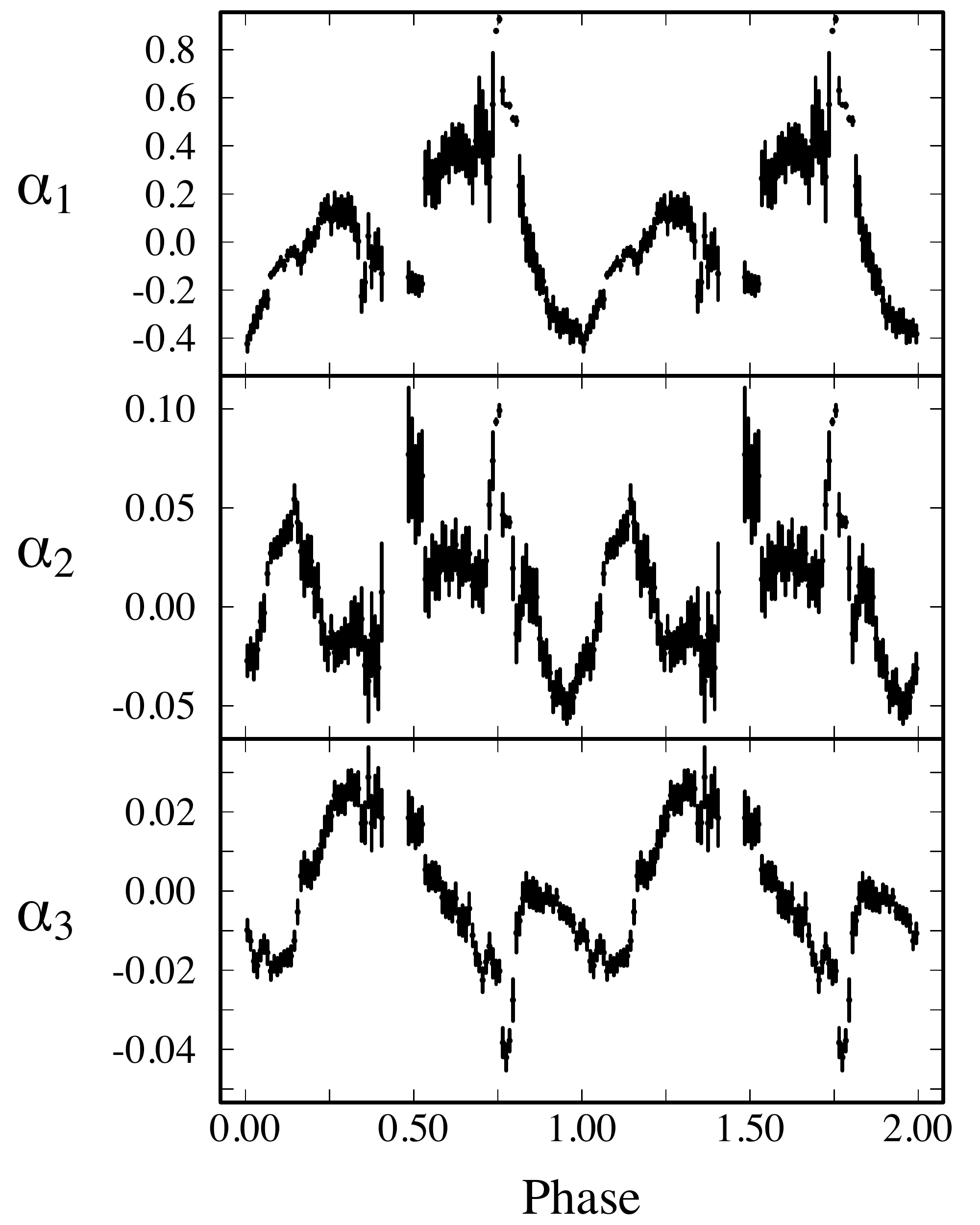}
\end{center}
\caption{The three first principal components phase-folded through all the \rxte\/ data.} \label{phasecomp}
\end{figure}

In addition to the analysis performed above, also ``colour'' (i.e principal component--component) and phase-folded principal component diagrams were constructed using the \rxte\/ data. These are plotted in Figs. \ref{compcomp} and \ref{phasecomp} respectively. Fig. \ref{compcomp} shows the first principal component as a function of the second principal component calculated from all the \rxte\/ pointings. We can see that in most observations the increase of both principal components happens more or less in a linear fashion, thus signalling a relationship between the two. Four observations (pointings 7, 8, 10 and 17) have been fitted with a linear regression and highlighted to show changes in the diagram. Also marked in Fig. \ref{compcomp} are the X-ray/radio states during the observations and a change from theÊFHXR state to the FIM state\footnote{The state nomenclature follows the classification of \citep{koljonen} where the FHXR and the FIM state correspond to intermediate-hard and intermediate-soft X-ray spectrum with radio flaring, respectively} can be seen approximately at the mean value (0.0) of the second principal component. As the radio flare fades, the slope of the fitted line decreases and the amount of variability in the second component increases. Fig. \ref{phasecomp} shows the principal components phase-folded to the orbit of Cyg X-3.

\subsubsection{\integral}      

Similarly to \swift\/ and \rxte\/, most of the variance (91\%) is caused by the first principal component. However, the LEV diagram (Fig. \ref{screeplot}, right panel) indicates that the components level off after the first component, and thus the \integral\/ data does not probe well the minor components. Despite this, we include the second component (contributing 3.5\% of the overall variance) in the analysis as it resembles the second principal component from \rxte\/, due to its similarity in shape with the variance spectrum (Fig. \ref{sigmaplot}), as well as in the variance fraction (Fig. \ref{pca}, lower middle row), the energy spectra (Fig. \ref{pca}, top and upper middle row), and exhibiting the same correlation behaviour with the X-ray flux (Fig. \ref{pca}, bottom row).
    
\subsection{Model of X-ray spectra} \label{model}

As mentioned in Section 1, with the aid of PCA the models of the X-ray spectra have to fulfil two requirements: fit the X-ray spectra with good statistics, as well as producing a comparable variation in the fit parameters (whatever they may be) matching the evolution of the principal components. In the following, we will go through the individual components of our best-fitting model for the X-ray spectra and explain our choices in detail. Since the bulk of the spectra were \rxte\/ spectra, we use them as a basis for comparison with the PCA results. Simultaneous fits with the \swift\/ and \integral\/ spectra were used to narrow down parameter ranges. All the data fitting was performed using \isis\/ (Interactive Spectral Interpretation System; \citealt{isis}).

\begin{figure}
\begin{center}
\includegraphics[width=0.5\textwidth]{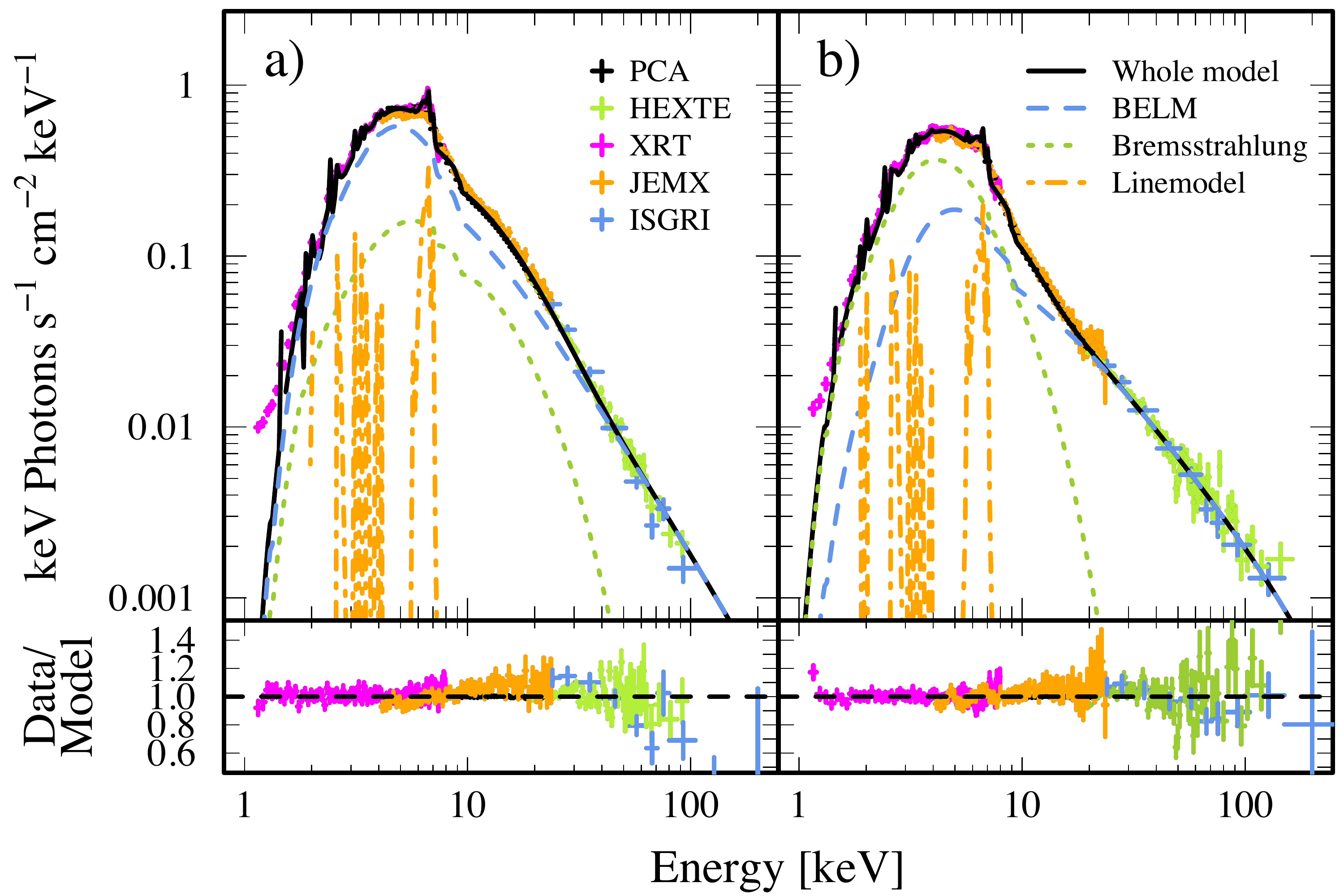}
\end{center}
\caption{An example of the FHXR (left panel) and FIM (right panel) energy spectrum with the best-fitting model (black, solid line) and the individual components (Comptonization: blue, dashed line; bremsstrahlung: green, dotted line; iron line: orange, dot-dashed line) overlaid and labelled.} \label{compspec}
\end{figure} 

\subsubsection{Absorption}

One of the crucial points to understanding the underlying physical properties of Cyg X-3 is its complex absorption profile. There is increasing evidence that the absorption is divided into multiple parts. One of the components is the absorption of the interstellar medium in the line-of-sight, which is estimated to be ~1.5 $\times 10^{22}$ cm$^{-2}$ \citep{dickey}, a result obtained using the 21 cm wavelength absorption. However, when studying the X-ray scattering halo \citep{predehl1} the absorption is found to be ~3.0 $\times 10^{22}$ cm$^{-2}$, clearly in disagreement with the former result. In addition, \citet{szostek2} studied the effect of the stellar wind on the emanating soft X-ray radiation from the compact object and found that the X-ray spectrum is well fitted when including cold clumps in the wind resulting in a partial covering factor of ~1\%. Therefore, it would be appropriate to use at least three different absorption components in the fits. However, as is the trend in modelling X-ray spectra, the number of free parameters in the models vastly increase when taking into account more physically realistic scenarios. A balance has to be found between a physically realistic model and a phenomenological model, but usually the former has so many variables that it will fit anything. Therefore, one has to strive for physically realistic models but endeavour to maintain the degrees of freedom as low as possible. To achieve this, we model the X-ray absorption with the simplest absorption profile, which consists of a uniform component (\phabs\/) and a partially covered component (\pcfabs\/). This model has successfully been applied to Cyg X-3 spectra previously in \citet{koljonen}, \citet{vilhu1}, \citet{szostek1} and \citet{hjalmarsdotter1}. We froze the parameter values according to fits performed with simultaneous \rxte\/, \integral\/ and \swift\/ spectra and let only the covering fraction vary when fitting only \rxte\/ spectra.  

\subsubsection{Line emission and absorption}

Cyg X-3 harbours a large number of emission lines that are especially prominent in the high-resolution energy spectra, e.g. \citet{paerels}. However, due to the low sensitivities of the \rxtepca\/ below $\sim$ 3.5 keV, only the iron line complex is important for the \rxte\/ spectra. \citet{paerels} showed that the iron line complex consists of helium-like and hydrogen-like iron ions (XXV/XXVI) at 6.7 keV and 6.9 keV, respectively, and cold iron K$\alpha$ line at 6.4 keV. However, these lines blend into a single broad iron line feature in the \rxtepca\/, and therefore we include only one gaussian line component in the model to represent this feature as a whole. More importantly, though, these ions also absorb hard X-ray photons producing helium-like and hydrogen-like iron edges in the energy spectra at 8.83 and 9.28~keV, which are fixed values in our model only their normalisation is allowed to vary in the fitting process. Due to the inclusion of \swift\/ spectra in simultaneous data fits we fit a line model by Savolainen et al. (in prep.) to fit the data. These lines most likely form in the stellar wind plasma surrounding the WR companion that are excited by the energetic radiation field from the vicinity of the compact object. However, the PCA results from \swift\/ (Section \ref{swift}) indicates that the iron lines could form in two locations where the other one contains more ionised iron.

\subsubsection{Continuum components}
      
Based on the PCA results we know that the X-ray variability is driven mainly by two components. From previous knowledge we know that the X-ray spectra of Cyg X-3 are well fitted by hybrid Comptonization (e.g. \citet{koljonen, hjalmarsdotter1}) and therefore we assume that the first principal component is also coupled to this model. For the Comptonization model we use a relatively new model, \belm\/ \citep{belmont}, that calculates self-consistently the equilibrium electron distribution taking into an account Compton scattering, synchrotron emission/absorption, pair production/annihilation and Coulomb collisions, assuming a spherical, magnetised and fully ionised proton-electron/positron plasma with radius $R$. Akin to the more widely used Comptonization model \eqpair\/ \citep{eqpair}, \belm\/ parametrizes energy injection as compactness 

\begin{equation}
l = \frac{\sigma_{T}}{m_{e}c^3} \frac{L}{R},
\end{equation}        

where $L$ is the power supplied to the plasma, $\sigma_{T}$ is the Thomson cross section and $m_{e}$ is the electron rest mass. The power can be supplied in three forms: 
(1) non-thermal electron acceleration ($l_{nth}$), with the injected electron distribution modelled as a power law (index $\Gamma_{inj}$, i.e. $n_{e}(\gamma)\propto\gamma^{\Gamma_{inj}}$) from energies $\gamma_{min}$ to $\gamma_{max}$, (2) thermal heating of electrons ($l_{th}$) e.g. in Coulomb collisions of hot ions, and (3) external radiation ($l_{s}$) from a geometrically thin accretion disk with a pure black body spectrum of temperature $kT_{bb}$. In a similar fashion to the electron-photon coupling, the magnetic energy density ($B^{2}/8\pi$) can be parametrized by 

\begin{equation}
l_{B} = \frac{\sigma_{T}}{m_{e}c^2} R \frac{B^2}{8\pi},
\end{equation}   

for tangled fields. The model spectra are tabulated for fitting in \isis\/ with fixed parameters to reduce the overall degeneracy of the parameters as well as computation time. The total compactness of the injected power that results in radiation ($l = l_{nth} + l_{th} + l_{s}$) can be estimated for Cyg X-3 assuming an average flux of $F_{0}$ = 1.3 $\times$ 10$^{-8}$ erg s$^{-1}$ cm$^{-2}$, a distance of 9 kpc, a black hole mass of 30 $M_{\odot}$, and a plasma size of 30 gravitational radii as:

\begin{equation}
l = 28 \Big(\frac{F}{F_{0}}\Big) \Big(\frac{d}{9 \mathrm{ kpc}}\Big)^2 \Big(\frac{30 M_{\odot}}{M}\Big) \Big(\frac{30 R_{G}}{R}\Big).
\end{equation}   

We consider only purely non-thermal electron injection (i.e. $l_{nth}\neq0$, $l_{th}=0$) with electron distribution parameters fixed at $\Gamma_{inj}=2$, $\gamma_{min}=1.3$ and $\gamma_{max}=1000$. We can therefore vary $l_{s}$ and derive the non-thermal compactness using $l = l_{nth} + l_{s}$. Since the electrons are allowed to thermalise through Coulomb collisions, this produces a hybrid thermal/non-thermal distribution in equilibrium. The hybrid model produced good results previously in \citet{hjalmarsdotter2} and references therein using \eqpair\/ and in \citet{koljonen} using \compps\/.

However, we find that not all the X-ray spectra studied are well fitted by using Comptonization alone, not even by allowing some of the fixed parameters mentioned above to vary as well. Also, as the PCA results show, a second principal component is required to explain all the X-ray variability and therefore we include a second continuum component in the model. As its overall effect in the PCA and X-ray spectra is smaller than the first principal component, i.e. Comptonization, degeneracies arise, and multiple components will fit the spectra. We found good fits when using reflection (\reflect, \citealt{reflect}), multicolour disc blackbody (\diskbb, \citealt{diskbb}), thermal bremsstrahlung (\bremss, \citealt{bremss}), or other thermal Comptonization models (\compst, \citealt{compst}; \comptt, \citealt{comptt}; \thcomp, \citealt{thcomp}), in addition to \belm\/. However, only a few of the above models produced the variability of the principal components, thus reducing the number of suitable models. Table \ref{modelcomparison} shows the chi-squared values of all the models together with correlations against the principal components and normalisation of the continuum components fitted to \rxte\/ pointings 1--13. The correlations were calculated using robust methods (e.g. \citealt{rousseeuw}) that offer good outlier detection, so that outliers in the data do not detract from the ``real'' correlation. Robust methods supply ways of detecting, weighing down (or rejecting altogether), and marking outliers that largely removes the need for manual screening. In addition to robust correlation we use robust regression, its coefficient of determination (R$^2$, essentially telling how close the data points cluster around the regression line) and the F-test probability of a null hypothesis in assistance to study the strength and significance of the correlation. The robust statistics are implemented using the \textsc{robust} library \citep{wang}.

Based on Table \ref{modelcomparison} we see that a lot of different setups will provide good reduced chi-squared values. However, when taking into account the correlations between the principal components and spectral model normalisations, four models (\diskbb, \bremss, \compst, and \comptt) stand out in Table \ref{modelcomparison}. All of these have optically thin ($\tau=0.5$) Comptonization and some thermal (albeit hot) component. These four models were then fitted to the whole dataset (bottom part of Table \ref{modelcomparison}). Based on these fits, the \compst\/ model performs the weakest and thus if the second principal component is due to thermal Comptonization it probably does not arise from cold seed photons. The \diskbb\/ model has slightly better correlation with the second principal component than \compst model. However, the disk temperature is very high ($\sim$ 5 keV), thus the radiation most likely is not coming from the disc. The possibility of a heated disc is discussed in Section \ref{thermalorigin}. Basically, the \bremss\/ and \comptt\/ models produce equally good fits and correlations. For the \comptt\/ model we set the seed photons to the same temperature as the BELM seed photons and the electron temperature was frozen to the values of the \diskbb\/ model. With these restrictions the optical depth for the thermal Comptonization is very high ($\tau \sim 5-10$), which, with an electron temperature of 5 keV, results in a Compton-y of $\sim$4 thus approaching saturated Comptonization. To summarise, the best-fitting model in spectral and variability terms has optically thin, rather thermal, Comptonization dominating the variability throughout the X-ray regime and a thermal, rather hot ($\sim$5 keV) plasma component producing variability in the $\sim$10--20 keV regime and radiating through Comptonization or bremsstrahlung. For the rest of the paper we have modelled the second spectral component as a thermal bremsstrahlung component (see discussion about the origin of the thermal component in Section \ref{thermalorigin}).

\begin{table*}
\caption{\rxte\/ spectral model comparison and corresponding statistical parameters. The model type is of the following (except for \reflect\/ which is a convolution model): \phabs\/ $\times$ \pcfabs\/ $\times$ \edge\/(1) $\times$ \edge\/(2) $\times$ (\egauss\/ + \belm\/ + X), where X is an additional component marked in the first column. The second column shows the mean reduced chi-square value and its standard deviation of the models fitted to the observations marked in the last column. The third column shows the robust correlation, the coefficient of determination (R$^{2}$) and the F-test probability of a null hypothesis for the normalisation parameter of \belm\/ with the first principal component. The fourth column is similar to the third except that the correlation statistics are taken from the normalisation parameter of the second spectral component in question (first column) with the second principal component.} \label{modelcomparison}
\begin{center}
\begin{tabular}{lllll}
\hline\hline
& & N$_{\rm{BELM}}$ vs. $\alpha_{1}$ & N$_{\rm{X}}$ vs. $\alpha_{2}$ & \\
Model X & $\bar{\chi}_{red}^{2}$ / $\sigma(\chi_{red}^{2}$) & Corr. / R$^{2}$ / Pr(F) & Corr. / R$^{2}$ / Pr(F) & Obs. N$^{\underline{o}}$\\
\hline
\diskbb(1)$^{\mathrm{a}}$ & 1.01/0.24 & 0.67/0.74/0.01 & 0.55/0.07/0.46 & 1--13 \\
\diskbb(2)$^{\mathrm{b}}$ & 0.94/0.16 & 0.88/0.73/1$\times$10$^{-5}$  & 0.91/0.85/6$\times$10$^{-3}$ & 1--13 \\
\thcomp(1)$^{\mathrm{c}}$ & 0.99/0.25 & 0.87/0.57/2$\times$10$^{-3}$  & -0.30/0.36/0.06 & 1--13 \\
\thcomp(2)$^{\mathrm{d}}$ & 0.99/0.25 & 0.95/0.68/8$\times$10$^{-5}$  & -0.36/0.33/0.09 & 1--13 \\
\thcomp(3)$^{\mathrm{e}}$ & 0.99/0.18 & 0.58/0.33/0.03 & -0.17/0.01/0.69 & 1--13 \\
\bremss & 0.92/0.18 & 0.86/0.68/2$\times$10$^{-4}$ & 0.95/0.75/3$\times$10$^{-6}$ & 1--13 \\
\compst$^{\mathrm{f}}$ & 0.93/0.17 & 0.81/0.59/3$\times$10$^{-4}$  & 0.88/0.59/4$\times$10$^{-4}$ & 1--13 \\
\comptt$^{\mathrm{g}}$ & 0.94/0.17  & 0.93/0.74/2$\times$10$^{-5}$  & 0.88/0.75/10$^{-6}$ & 1--13 \\
\reflect$^{\mathrm{h}}$ & 0.90/0.18 & 0.86/0.70/3$\times$10$^{-5}$ & 0.38/0.13/0.24 & 1--13 \\
\hline
\diskbb(2) & 0.99/0.17 & 0.94/0.87/2$\times$10$^{-16}$ & 0.64/0.63/3$\times$10$^{-3}$ & 1--27 \\
\bremss & 1.02/0.20 & 0.97/0.89/7$\times$10$^{-15}$ & 0.83/0.70/7$\times$10$^{-8}$ & 1--27 \\
\compst & 0.99/0.17 & 0.95/0.83/5$\times$10$^{-14}$ & 0.52/0.39/0.01 & 1--27 \\
\comptt & 1.00/0.17 & 0.94/0.87/3$\times$10$^{-15}$ & 0.76/0.68/5$\times$10$^{-8}$ & 1--27 \\
\hline
\end{tabular}
\end{center}
\begin{list}{}{}
\item[$^{\mathrm{a}}$] Colour temperature is set equal to the temperature of the seed photons for \belm.
\item[$^{\mathrm{b}}$] Colour temperature thawed.
\item[$^{\mathrm{c}}$] Seed photon temperature is set equal to the temperature of the seed photons for \belm\/ and the electron $\Gamma$ is frozen at 2.
\item[$^{\mathrm{d}}$] The same as above, but with seed temperature frozen to 50 eV.
\item[$^{\mathrm{e}}$] The same as above, but with electron $\Gamma$ thawed.
\item[$^{\mathrm{f}}$] Electron temperature is equal to colour temperature in model \diskbb(2).
\item[$^{\mathrm{f}}$] Seed photon temperature is set equal to the temperature of the seed photons for \belm\/, electron temperature is set equal to colour temperature in model \diskbb(2), and spherical approximation scheme.
\item[$^{\mathrm{h}}$] Iron abundances, metal abundances and cosine of inclination set to 1.48, 1.58 and 0.5 respectively.
\end{list}
\end{table*}

We would also like to note that a marginal effect ($\sim$1\%) on the variance spectrum is caused by the third principal component which, based on the shape in the variance spectrum, is most likely the reflection component from the accretion disc. As the reflection bump would cause only a very minor adjustment to the X-ray spectra above 10 keV we chose not to include it in the model, and we consider the iron K$\alpha$ line from reflection to be blended in the gaussian line of the model. However, as discussed in the next section, the centroid of the gaussian line follows the third principal component so that when the centroid approaches 6.4 keV (i.e. the K$\alpha$ line) the proportion of variability of the third principal component increases and vice versa. This evolution also gives credibility to the interpretation that the third principal component is caused by reflection. 

The best-fitting models for all \rxte\/ spectra are tabulated in Table \ref{wholemodel1}. The best-fitting models for simultaneous \rxte\/, \integral\/ and \swift\/ spectra are tabulated in Table \ref{wholemodel2} with two examples from the FHXR and FIM states plotted in Fig. \ref{compspec} together with individual model components. 
The implications of these results are discussed in Section \ref{discussion}.

\subsection{Simulating the effect of parameter variability on the variance spectra} \label{simulations}

\begin{figure*}
\begin{center}
\includegraphics[width=1.0\textwidth]{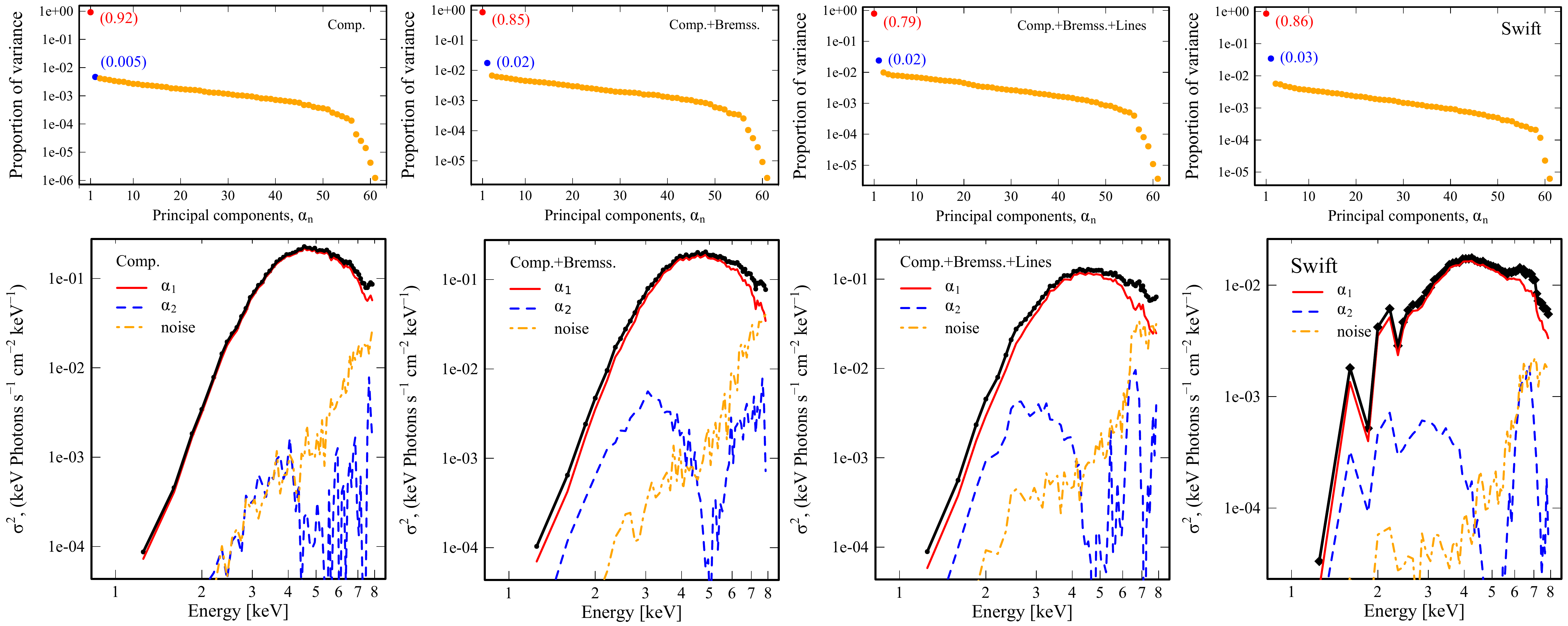}
\end{center}
\caption{Simulated variance spectra (bottom panels) and LEV diagrams (top panels) for \swift\/. See text for details. The first panels from the left show the LEV diagram and the variance spectrum when only the \belm\/ normalization was varying. The second panel from the left shows the LEV diagram and the variance spectrum when \belm\/ and \bremss\/ normalisations were varying. The third panel from the left shows the LEV diagram and the variance spectrum when all the component normalisations were allowed to vary, and the last panel shows the actual LEV diagram and the variance spectrum calculated from the data.} \label{swiftsimulation}
\end{figure*}

\begin{figure*}
\begin{center}
\includegraphics[width=1.0\textwidth]{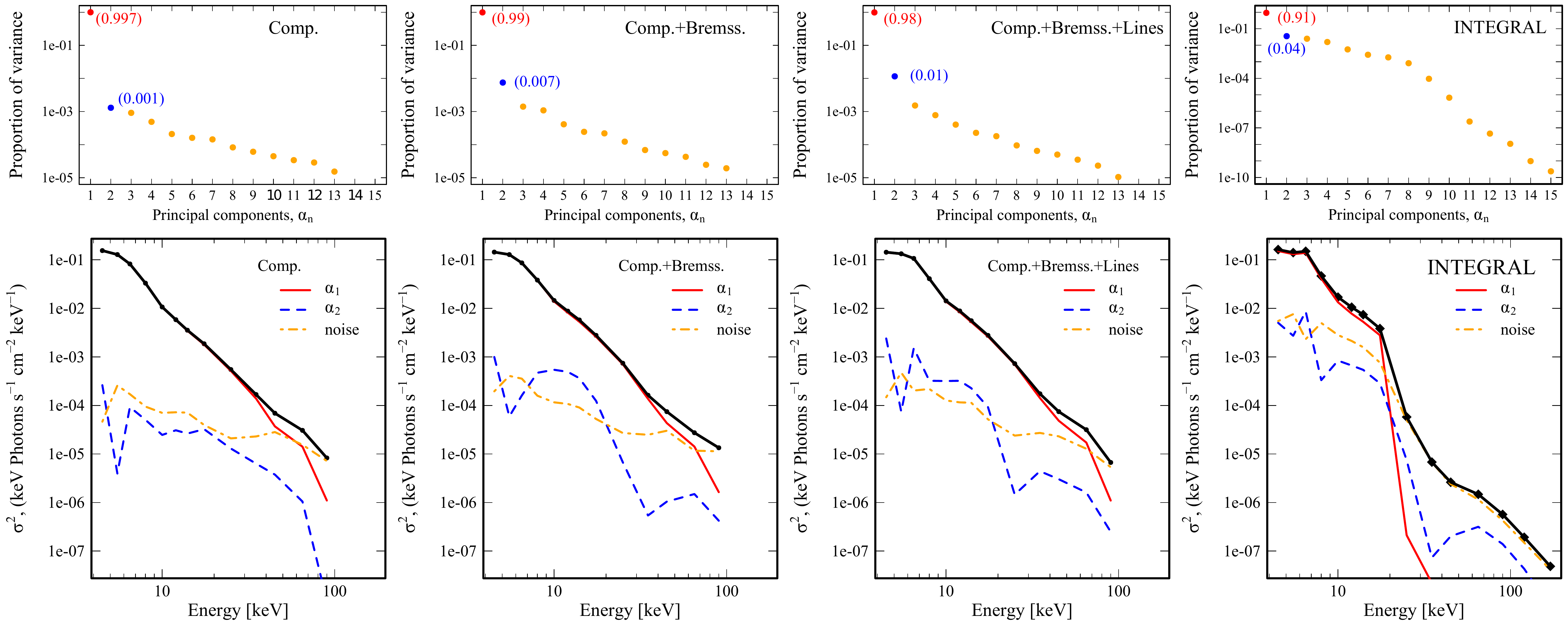}
\end{center}
\caption{The same as in Fig. \ref{swiftsimulation} but for \integral\/ data.} \label{integralsimulation}
\end{figure*}

\begin{figure*}
\begin{center}
\includegraphics[width=1.0\textwidth]{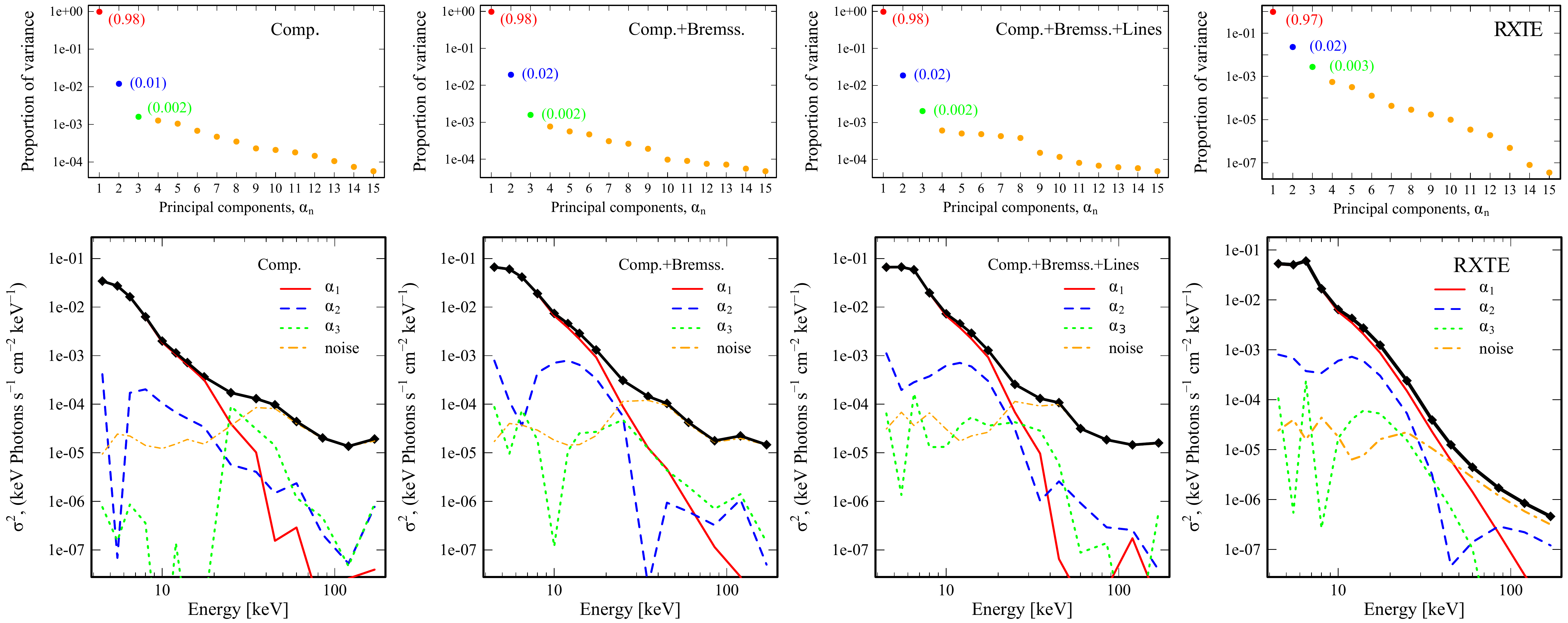}
\end{center}
\caption{The same as in Fig. \ref{swiftsimulation} but for \rxte\/ data.} \label{rxtesimulation}
\end{figure*}

To double-check that the above best-fitting model indeed reproduces the PCA results presented in Section \ref{pcaresults}, we have simulated lightcurve-spectra for \swift\/, \rxte\/ and \integral\/ data to see how the variability of the spectral parameters affects the variance spectrum. This simulation also serves as a method to check the effects of \swiftxrt\/ instrument response as mentioned in Section \ref{swift}. The method for creating the lightcurve-spectra is as follows: the auxiliary response files and redistribution matrix files of a pointing were fed into \isis\/ together with the best-fitting model. To reproduce the variability reminiscent of that seen in the original lightcurves, the normalization values of \belm, \bremss\/ and emission lines were multiplied by the values of the corresponding principal components. This step also allows to switch between different setups, e.g. first allowing only the normalization of one component to vary while keeping the others at zero and after that adding more components. Then the spectrum is faked according to the same exposure times that were used in the actual lightcurves (see Section \ref{observations}). After faking the fluxes from the same energy bands that were used in the original analysis were retrieved. This procedure is then looped through to obtain the same length for the simulated lightcurves as in the original data (with varying random seed for the {\tt fakeit} function). The results are depicted in Figs. \ref{swiftsimulation}, \ref{integralsimulation} and \ref{rxtesimulation}. In these figures the first panel from the left shows the variance spectrum (bottom) and LEV diagram (top) when only \belm\/ was varying, the second panel shows when \belm\/ + \bremss\/ were varying, the third panel shows when \belm\/ + \bremss\/ + emission lines were varying, and the last panel shows the original variance spectrum and LEV diagram. 

The simulated data of \swift\/ and \integral\/ show that the second component is a mixture of bremsstrahlung and line emission, which could suggest a reflection model. However, the simulated (and original) \rxte\/ PCA data (which is the best data set) show that this thermal (bremsstrahlung) component is distinct from the reflection component. Neither do the model fits support the reflection component. In addition, the whole X-ray spectrum is rather thermal during the major flare (the hard tail rises later on), so there are not too many hard photons available for producing a strong reflection component.

\section{Discussion} \label{discussion}

Once successful fits to the spectra have been generated, we can examine in more detail what the results imply in terms of parameter evolution as the major flare decays, as well as how the parameters vary along the binary orbit.

\subsection{Parameter correlations}

\begin{figure}
\begin{center}
\includegraphics[width=0.5\textwidth]{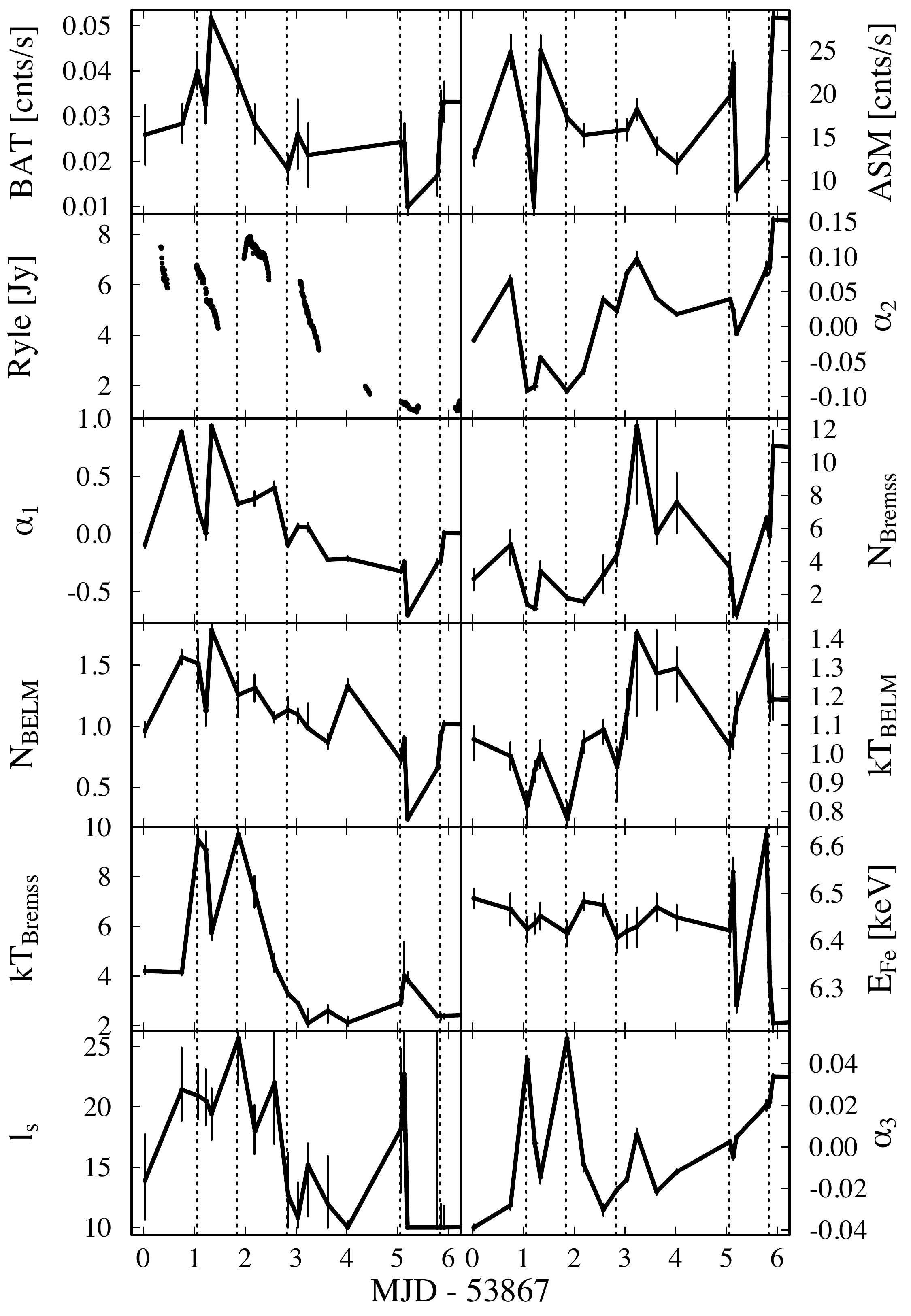}
\end{center}
\caption{Lightcurves of \rxteasm\/, \swiftbat\/, Ryle, selected fit parameters and three principal components during the May outburst. The vertical dotted lines mark observations (pointings 3, 6, 9, 14 and 18) where the seed temperature for Comptonization expresses a local minimum. These times correspond somewhat to local minima in the second principal component, bremsstrahlung normalization, gaussian line centroid and to maxima of soft photon compactness for Comptonization, and are attributed to phase interval 0.2--0.4. See text for implications.} \label{bremsprlc}
\end{figure}

\begin{figure}
\begin{center}
\includegraphics[width=0.5\textwidth]{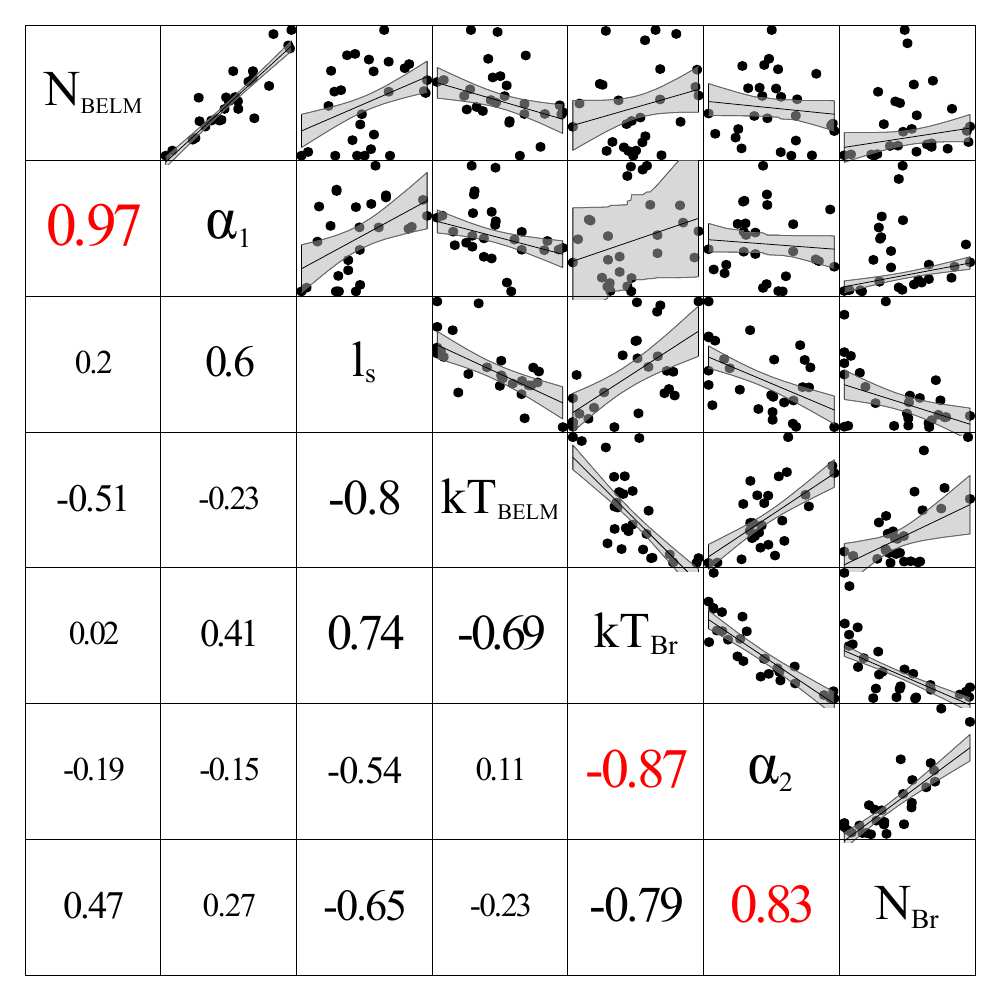}
\end{center}
\caption{The scatter plot matrix of the selected fit parameters and first two principal components with robust correlations (e.g. \citealt{rousseeuw}) drawn and written in appropriate grid cells. The black lines show the best linear robust regression and the grey ribbons show the standard 95 \% significance limits. The red numbers correspond to the R-squared value (i.e. the goodness of the robust fit or the correlation between the predicted values with the actual values.) over 0.6. Thus, coloured numbers close to $\pm$1 imply correlation.} \label{bremscor}
\end{figure}

As required, the normalisations of the different spectral components follow the principal components (see Figs. \ref{bremsprlc} and \ref{bremscor}), i.e. the first principal component $\alpha_{1}$ tracks the normalization of the Comptonization model (\belm) and the second principal component $\alpha_{2}$ the normalization of the bremsstrahlung model (\bremss). The third principal component $\alpha_{3}$ most likely corresponds to a reflection model (not included in the fits) as was discussed in Section \ref{results}. Other interesting possibly correlated parameters include the Comptonized soft seed temperature ($kT_{bb,comp}$) that appears to track the bremsstrahlung normalization, which could indicate that most of the seed photons for Comptonization are produced by bremsstrahlung emission. This interpretation is reinforced by the linear relationship seen between the first and the second principal components in Fig. \ref{compcomp}, and also by the convergence of the bremsstrahlung temperature ($kT_{bb,brems}$) and $kT_{bb,comp}$ in Fig. \ref{bremscor}. However, there are local minima in the $kT_{bb,comp}$ that are marked with dashed vertical lines in Fig. \ref{bremsprlc}. At the same time local minima are found (in most cases) for $\alpha_{2}$, the bremsstrahlung normalization, and the gaussian line centroid, while the external seed photon compactness and $\alpha_{3}$ (reflection) reaches local maxima. These times correspond to phase interval 0.2--0.4 (when the increased variability and low-frequency QPOs were identified by \citealt{koljonen2}) and one can also see that the gaussian centroid approaches 6.4~keV, indicating increasing cold iron K$\alpha$ emission. Thus, it is reasonable to assume that this is when the impact of the accretion disc is more prominent. Based on these results we can reinforce the conclusion in \citet{zdziarski} that there is a thermal plasma component in the system in addition to the accretion disk/corona combination. It is also evident that the phase interval 0.2--0.4 is somehow distinct, and it is in this phase that we expect to probe deeper into the system. 

\subsection{Orbital variations}

\begin{figure}
\begin{center}
\includegraphics[width=0.5\textwidth]{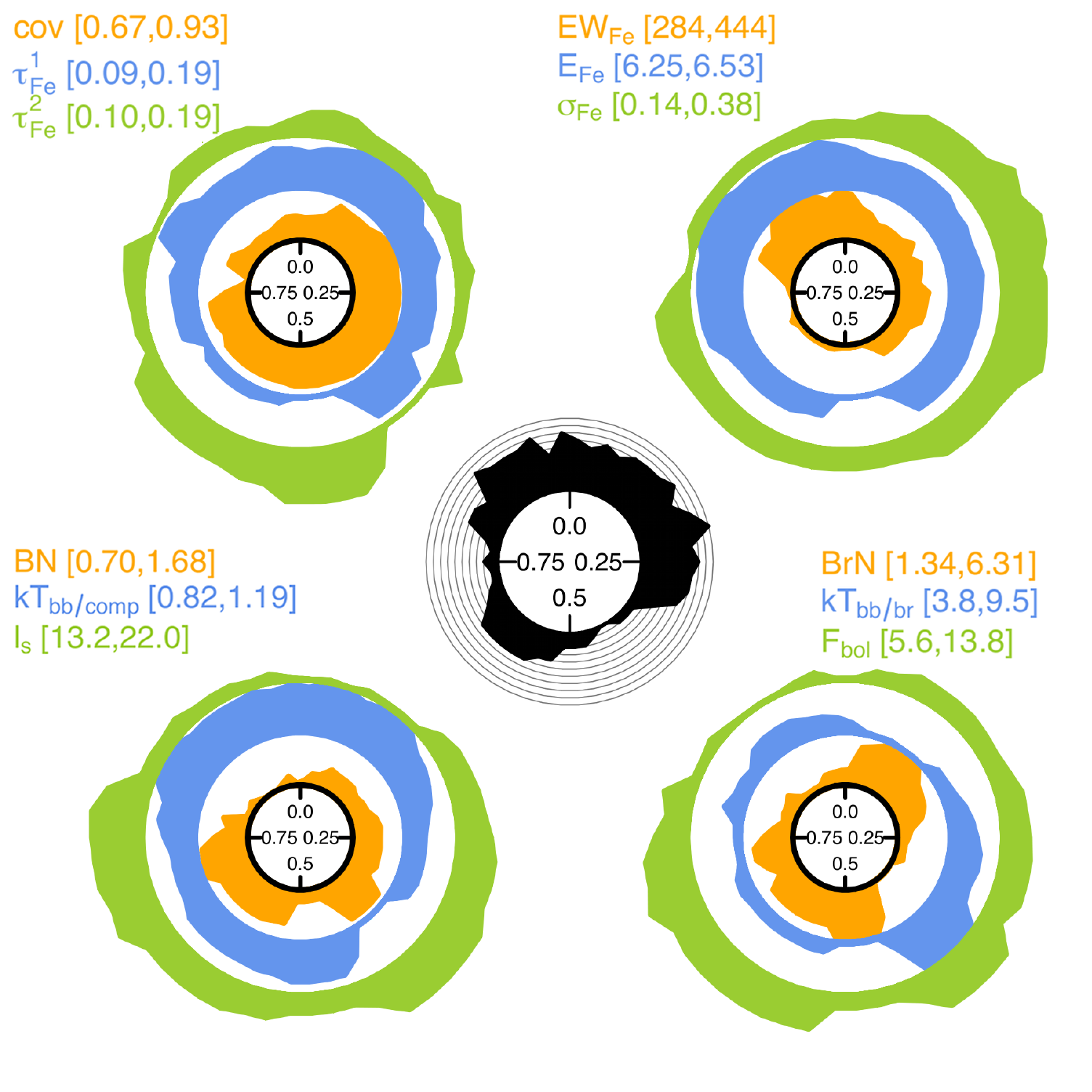}
\end{center}
\caption{Phase-circle diagrams displaying the phase variation of fit parameters and bolometric X-ray flux (3.5-200 keV). Each of the four coloured diagrams present three fit parameters plotted on a circle, where the angle shows the orbital phase and the height of the curve shows the mean parameter value from stacked observations (the number of observations shown at the center diagram for each phase bin) with the minimum value on the inner circle and the maximum value on the outer circle. In the brackets are the minimum and maximum values of a given parameter. The individual fits with parameters and their units can be found in Table \ref{wholemodel1}. \textit{Upper left}: absorption parameters: the covering fraction of the partial absorber ($cov$) and the optical depths of the absorption edges of H- and He-like iron ($\tau^{1}_{Fe}$ and $\tau^{2}_{Fe}$). \textit{Upper right}: iron line complex parameters: Equivalent width ($EW_{Fe}$), centroid ($E_{Fe}$) and line width ($\sigma_{Fe}$). \textit{Lower left}: Comptonization parameters: normalization ($BN$), soft seed photon temperature ($kT_{bb/comp}$) and the compactness of the radiation from soft seed photons ($l_{s}$). \textit{Lower right}: bremsstrahlung parameters (normalization $BrN$ and plasma temperature $kT_{bb/br}$) and bolometric X-ray flux ($F_{bol}$.) \textit{Center}: the number of pointings contributing to a given phase bin. The concentric circles show the number of observations from 0 (smallest circle) to 10 (largest circle).} \label{phasecircle}
\end{figure}

Since Cyg X-3 has a very short orbital period, it is possible to investigate possible changes in the X-ray spectra as a function of phase and relate them to geometrical properties of the source. In order to look for phase-dependent changes in the emission components acquired from the \rxte\/ spectral fits, we have stacked the values of the fit parameters from each observation into so-called {\it phase-circle} diagrams (Fig. \ref{phasecircle}). Each of these diagrams displays three fit parameters plotted in concentric rings, where the angle shows the orbital phase of the system and the width of the curve shows the mean parameter value from the stacked observations with the minimum value on the inner circle and the maximum value on the outer circle of the corresponding ring. The phase-circle diagram allows a quick assessment of how the fit parameters vary along the orbit and how they relate to each other. Based on Fig. \ref{phasecircle}, we see the following: 

(i) The covering fraction of the partial absorber has a minimum at phase interval 0.75--0.00. During this phase interval the iron line centroid and the equivalent width approach maximum values and the line width approaches minimum value. Thus, during this phase we see a region which is not as strongly absorbed and produces a strong iron line that is highly ionised but narrow. A possible interpretation could be that the compact object has swept up the wind and hence is at least temporally in a lower density region; 

(ii) The normalization of Comptonized emission and the external photon field compactness follow the overall phase variation of the bolometric X-ray flux. This is in line with the notion that almost all the X-ray flux observed comes from the Comptonized emission of external photon field that in turn photoionises a region producing an iron line. The optical depth of the H-like iron edge and the equivalent width of the iron line show anti-correlated behaviour to the X-ray flux so that the maximum occurs around phase 0.0 and minimum at phase 0.5. This is most likely due to increasing stellar wind depth along the orbit;

(iii) During phase interval 0.3--0.4 the normalization of the bremsstrahlung emission and the temperature of the soft seed photons undergoing Comptonization is at minimum. This might be an effect of the dearth of available data. However, the phase interval 0.2--0.4 also corresponds to the epochs marked with dotted lines in Fig. \ref{bremsprlc}, where the centroid of the gaussian line complex dips closer to 6.4 keV. It is reasonable to assume that during this phase the geometry of the system changes, e.g. the disc photons are able to see the Comptonizing component, which is also supported by the increase in variability and the presence of low-frequency QPOs during this phase \citep{koljonen2};

(iv) The normalization of bremsstrahlung emission, which is proportional to the electron and ion densities and the volume of the emission region, has two maxima at phases 0.0--0.25 and 0.5--0.75. This also explains why the second principal component as seen in \rxte\/ and \integral\/ data is anticorrelated with the X-ray fluxes Fig. \ref{pca}. Therefore, either the electron and ion densities or the volume (or both) increase at these phase intervals. If these regions are associated with the interaction sites between the compact object and the stellar wind this might imply that the stellar wind is restricted to these phases, e.g. it is a disk-like structure around the WR star as proposed in \citet{fender}. 
  
For the \comptt\/ model the Comptonization normalization behaves exactly like the bremsstrahlung normalization. The only difference with the bremsstrahlung model comes from the inclusion of optical depth in \comptt\/ which produces a maxima around phase 0.75. 

Based on the phase-folded principal components (see Fig. \ref{phasecomp}), we see that the first principal component has maxima between phases 0.2--0.4 and 0.6--0.8, much in line with the \belm\/ normalization. However, a sharp peak is present at phase 0.75. A similar peak is observed also in the second principal component with additional maxima between phases 0.0--0.2 and around phase 0.5. 

\subsection{Location of the Comptonizing region}

X-ray/radio state changes seem to be driven by the interplay of two spectral components, Comptonization and bremsstrahlung, during a major radio flaring episode. From the definition of \citet{koljonen} the change from the FHXR state to the FIM state signals softening of the X-ray spectra. This is coupled to a decrease of the radio flux density as the major radio flare fades (see Fig. \ref{obs} bottom left panel and Table \ref{xobs}). The Compton upscattering seem to be more efficient during the FHXR state. Fig. \ref{compcomp} shows a stronger increase in the first principal component (which is coupled to the Comptonization normalization) during this state than for the second principal component (which is coupled to the bremsstrahlung normalization). Also, Fig. \ref{compspec} shows that the X-ray flux attributed to Comptonization decreases and the X-ray flux attributed to bremsstrahlung increases when the source moves from the FHXR state to the FIM state. Possible interpretations of this behaviour are: (1) the Comptonizing region is located in the jet and it is moving away from the pool of seed photons. Close to a major radio flare peak, most of the seed photons see the Comptonizing region and produce a harder spectra. When the Comptonizing region moves away and/or dilutes as the major radio flare fades more seed photons arrive to us unscattered. (2) The increase of the seed photons (possibly corresponding to the increase of the stellar wind) cools the Comptonization region at the base of the jet so that there is a decrease in Comptonization and increase in unscattered seed photons. This would also serve as a mechanism to quench the jet ejection episode. In essence these two scenarios can be dubbed as the distant location and the close location of the Comptonizing region, respectively. 

While it is possible that the high energy electron population originates close to the compact object it seems more probable that it originates further out, e.g. in the jet, because of the low optical depth for the Comptonized emission. When using the above best-fitting model, we found that the Thomson optical depth for the Comptonization approaches its lowest possible value (in our model this corresponds to $\tau_{T}$=0.5). A close origin is also not favourable if the system inhabits a WR companion star, as its dense stellar wind most likely enshrouds the compact object causing extra scattering of the photons originating close to the compact object.

\subsection{Comparison to other microquasar systems}

In addition to Cyg X-3, similar thermal, hot components have been identified in other microquasar/XRB systems. \citet{titarchuk} found that eight IS/HSS spectra from GRS 1915$+$105 required a ``high-temperature black body-like'' profile; these were somewhat related to epochs of radio activity. Also, \citet{mineo} found that spectra in the ``heartbeat" state of GRS 1915$+$105 can be fitted with models including a black body component with colour temperature 3--6 keV. \citet{seifina} found 24 IS spectra from SS 433 during radio outburst decay that had a strong black body-like component with colour temperature 4--5 keV. In addition, the spectra from the black hole soft X-ray transients GS 2000$+$25, GS 1124$-$68 and XTE J1550$-$564 were successfully fitted with an additional thermal Comptonization component with colour temperature 2--4 keV and high optical depth $\tau \sim 5$ \citep{zycki}. This prompts the question: is this thermal component something intrinsic to black hole systems/microquasars? Is the emission mechanism bremsstrahlung or thermal Comptonization as it appears to be in Cyg X-3 or something else?

\subsection{Origin of the thermal component} \label{thermalorigin}

In the papers mentioned above a multitude of different origins were presented for the thermal component. Here we briefly review them and consider a few extra options as well.  Since rigorous testing of the individual origin is beyond the scope of this paper, this discussion will be somewhat minimal.

\textbf{Gravitationally redshifted annihilation line?} 
In \citet{titarchuk} the thermal component was interpreted as a highly gravitationally redshifted annihilation line. However, a clear view close to compact object would also result in a better view of the accretion disc which is not observed, based on our modelling, at the same time with the thermal component. Rather, it seems that the impact of the disc is more prominent when the thermal component is at minimum.

\textbf{Hot disc?} 
One option is that the accretion disc is partly heated by the hard X-rays (thus reaching colour temperatures higher than the usual $\sim$ 1 keV). Physically, this would represent e.g. a patchy corona above the disk with hard X-rays arising from magnetic reconnection. However, based on the fluxes this scenario is not favourable. The hard X-ray flux ranges between $F_{10-200 \mathrm{keV}} \sim 0.7-6.3$ (in units of $10^{-9}$ erg cm$^{-2}$ s$^{-1}$) while the thermal component flux (as modelled by thermal bremsstrahlung) ranges between $F_{th} \sim 1.0-7.9$ with $F_{th} > F_{10-200 \mathrm{keV}}$ in nearly all \rxte\/ pointings. Also, we find the reflection component to be very small or non-existent and thus, reprocessed disc emission is not likely the origin of the thermal component.

\textbf{Three-phased accretion?} 
This scenario is based on the idea \citep{zycki} that a transitionary warm plasma region could form between the cold accretion disc and the hot inner flow/corona. This would then supply seed photons for Comptonization in the corona. However, it is hard to envision how the orbital changes are incorporated into this scenario. 

\textbf{Thermal jet?} 
In a similar case to the previous point, \citet{memola} studied a thermal plasma component in the jet. According to this study the innermost jet can reach temperatures above 1~keV up to 1~MeV, thus suggesting that thermal X-ray emission might originate from this region. The observations of the thermal components during intermediate states, during which strong jets are observed (especially for Cyg X-3), supports this scenario. However, the evolution of the principal components shows that while the major radio flare decays the thermal component becomes more prominent, which would point to a non-jet origin. 

\textbf{Hybrid electron distribution?} 
By using \belm\/ we assume a hybrid electron distribution in the coronal electrons {\it ab initio} motivated by the successful use of hybrid models in fitting Cyg X-3 spectra (e.g. \citealt{koljonen} for \compps\/). Interestingly, \compps\/ was found to successfully fit spectra of GRS 1915$+$105 \citep{mineo}, XTE J1550$-$564 and some from GS 1124$-$68 \citep{zycki}. However, as portrayed by the PCA results and the unsuccessful spectral fits using just \belm\/, at least for Cyg X-3, the need for additional thermal component is evident.   

\textbf{Scattering cloud?} 
\citet{zdziarski} argued for the presence of a Thomson-thick, low-temperature plasma cloud surrounding the compact object in Cyg X-3 in order to explain the lack of high frequencies in the power spectra and the peculiar hard state X-ray spectra with $\sim$ 30 keV cut-off by Compton downscattering. This cloud could be the result of the collision between the WR stellar wind and the compact object. The plasma parameters found in \citet{zdziarski} were $kT\sim3$ keV and $\tau\sim 7$. For these parameters the thermal bremsstrahlung emission becomes a substantial source for photons which get amplified by Comptonization in the plasma cloud. The orbital changes in the thermal component could arise if the wind is asymmetric and restricted to certain phases along the orbit.  

\subsection{Synopsis}

\begin{figure*}
\begin{center}
\includegraphics[width=1.0\textwidth]{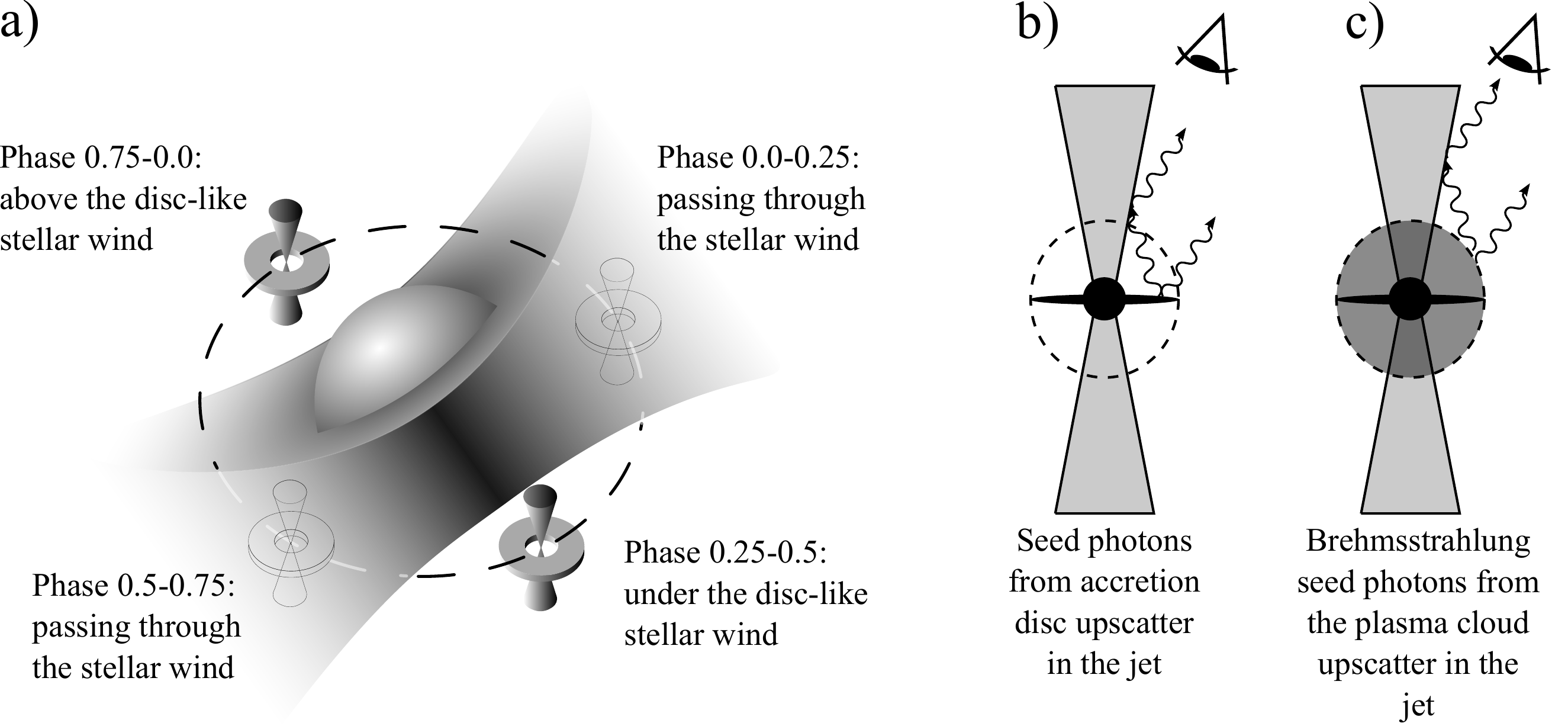}
\end{center}
\caption{\textit{a)} A sketch depicting the geometry of the system, with the Wolf-Rayet companion surrounded by a disc-like stellar wind and the companion object orbiting it (shown in four different orbital phases). Depending on the orbital phase the compact object is either inside or outside the stellar wind. \textit{b)} The two different soft seed populations arise from the accretion disc and \textit{c)} from the interaction of the compact object with the stellar wind.} \label{sketch}
\end{figure*}

Based on the above discussion we propose a scenario which explains the observed behaviour of the system (see Fig. \ref{sketch}):

(i) The principal emission process responsible for most of the variability in the X-rays during jet ejection events is the Compton scattering of soft seed photons from two distinct thermal populations by the high-energy electrons confined most likely in shocks in the relativistic jet. 

(ii) One source of these thermal populations is a thermal plasma cloud around the compact object, but which is restricted to phases 0.0--0.25 and 0.5--0.75. We interpret this (inspired by \citealt{zdziarski}) as an interaction between the compact object and the stellar wind. Therefore the stellar wind could be formed around the WR star as a disc. This is in line with the studies that show that approximately 15\% of WR stars have anisotropic winds, which are most likely caused by equatorial density enhancements produced by high rotation rates \citep{harries}. In this plasma cloud the bremsstrahlung emission process becomes important and provides soft seed photon population that Compton upscatter in the jet. 

(iii) Outside the disc-like wind (provided that the compact object does not orbit the WR star exactly in the plane of the wind) this interaction is much weaker. The disc-like wind provides higher absorption when the compact object is under the disc along the line of sight (phase 0.25--0.5) and lower absorption when above the disc (phase 0.75--0.0). However, this is more a matter of going from a high density to a lower density part of the wind. But even in the low density part of the wind the density still needs to be rather high in order to wipe out the timing signature above 0.1 Hz \citep{koljonen2,axelsson}.   

(iv) During times outside the disc-like wind, a second source of soft seed photons arises from the accretion disc. Hints for glimpsing the underlying disc is obtained by the increasing X-ray variability during these phases. \citet{koljonen2} found that during phases $\sim$0.3 and $\sim$0.7 the X-ray variability peaked above what could be expected based on the orbital modulation. Also, on two occasions a mHz QPO was detected around phase $\sim$0.3. In addition, based on the spectral fits on the \rxte\/ data the phase interval 0.3--0.4 shows a minimum in the soft seed photon temperature as well as the centroid of the gaussian line complex approaching 6.4~keV.

\section{Conclusions} \label{conclusions}

In this paper we have studied in detail the X-ray behaviour of Cyg X-3 during a major radio flaring event. We have shown that additional information is obtained by using Principal Component Analysis on fast X-ray spectral data ($\sim$ 1 minute), which helps to discriminate between different spectral models fitted to the time-averaged spectra ($\sim$ 3~ks). 
We have found that most of the variability is caused by two principal components whose evolution based on spectral fits is best reproduced by a Comptonization component and a bremsstrahlung component that most likely arises from the interaction between the compact object and the stellar wind from the Wolf-Rayet companion. Following the fit parameter evolution over the orbit, it appears that the bremsstrahlung component is restricted to certain phases that are opposite to each other, which could indicate that the stellar wind is formed as a disc-like wind around the companion star. Therefore, during orbital phases when the compact object is outside the wind we get a glimpse deeper into the directly of the accretion disc. This is also supported by the findings of increased variability and QPOs in the X-ray lightcurves by \citet{koljonen2}. Thus, we propose a scenario for Cyg X-3 where the soft seed photon population arises from two different origins: from bremsstrahlung and accretion disc that in turn get Comptonized further down in the jet.

\section*{Acknowledgbesments}

KIIK gratefully acknowledges support from the Finnish graduate school of astronomy and space sciences, a grant from Jenny ja Antti Wihurin s\"a\"ati\"o and Acadamy of Finland grant (project num. 125189). This research has made use of data obtained from the High Energy Astrophysics Science Archive Research center (HEASARC), provided by NASA's Goddard Space Flight center.

\appendix

\section[]{Fit parameters}

\begin{table*}
\caption{\texttt{phabs} $\times$ \texttt{pcfabs} $\times$ \texttt{edge(1)} $\times$ \texttt{edge(2)}  $\times$ (\texttt{gauss} + \texttt{belm} + \texttt{bremss})} \label{wholemodel1}
\begin{center}
\begin{tabular}{lcccccccc}
\hline\hline
Component & Parameter & Unit & N$^{\underline{o}}$ 1 & N$^{\underline{o}}$ 2 & N$^{\underline{o}}$ 3 & N$^{\underline{o}}$ 4 & N$^{\underline{o}}$ 5 & N$^{\underline{o}}$ 6 \\
\hline
\texttt{phabs} & nH$_{0}$ & $10^{22}$ & 			$3f$ 							&	$3f$					&	$3f$					&											$3f$							&	$3f$					&	$3f$					\\
\texttt{pcfabs} & nH$_{1}$ & $10^{22}$ & 			$12f$						&	$12f$				&	$12f$				&											$12f$						&	$12f$				&	$12f$				\\
& cov & $$ &								$0.81^{+0.04}_{-0.04}$			&	$0.86^{+0.03}_{-0.04}$	&	$0.89^{+0.05}_{-0.04}$	&											$0.82^{+0.03}_{-0.03}$			&	$0.89^{+0.05}_{-0.04}$	&	$0.64^{+0.04}_{-0.04}$	\\
\texttt{edge(1)} & E$_{\rm{Fe}}^{\rm{ion}}$ & keV &	$9.28f$						&	$9.28f$				&	$9.28f$				&											$9.28f$						&	$9.28f$				&	$9.28f$				\\
& $\tau(1)_{\rm{Fe}}^{\rm{ion}}$ & $$ & 			$0.17^{+0.04}_{-0.04}$			&	$0.13^{+0.04}_{-0.04}$	&	$0.17^{+0.03}_{-0.04}$	&											$0.13^{+0.03}_{-0.04}$			&	$0.17^{+0.03}_{-0.04}$	&	$0.12^{+0.04}_{-0.04}$	\\
\texttt{edge(2)} & E$_{\rm{Fe}}^{\rm{ion}}$ & keV &	$8.83f$						&	$8.83f$				&	$8.83f$				&											$8.83f$						&	$8.83f$				&	$8.83f$				\\
& $\tau(2)_{\rm{Fe}}^{\rm{ion}}$ & $$ & 			$0.14^{+0.04}_{-0.04}$			&	$0.12^{+0.04}_{-0.04}$	&	$0.10^{+0.04}_{-0.03}$	&											$0.14^{+0.04}_{-0.03}$			&	$0.10^{+0.04}_{-0.03}$	&	$0.14^{+0.04}_{-0.04}$	\\	
\texttt{gauss[Fe]} & EW$_{\rm{Fe}}$ & eV & 		$393^{+28}_{-24}$			&	$285^{+33}_{-30}$			&	$344^{+30}_{29}$		&											$382^{+26}_{-24}$				&	$299^{+28}_{-25}$		&	$367^{+43}_{-29}$\\
& E$_{\rm{Fe}}$ & keV &						$6.49^{+0.02}_{-0.02}$			&	$6.47^{+0.03}_{-0.03}$	&	$6.43^{+0.03}_{-0.03}$	&											$6.44^{+0.02}_{-0.02}$			&	$6.43^{+0.03}_{-0.03}$	&	$6.48^{+0.02}_{-0.02}$	\\
& $\sigma_{\rm{Fe}}$ & keV & 					$0.24^{+0.05}_{-0.05}$			&	$0.28^{+0.08}_{-0.09}$	&	$0.29^{+0.06}_{-0.06}$	&											$0.24^{+0.05}_{-0.06}$			&	$0.29^{+0.06}_{-0.06}$	&	$0.14^{+0.06}_{-0.14}$	\\
\texttt{BELM} & norm & $$ &					$0.96^{+0.07}_{-0.05}$			&	$1.56^{+0.06}_{-0.06}$	&	$1.51^{+0.2}_{-0.2}$		&											$1.13^{+0.12}_{-0.13}$			&	$1.51^{+0.20}_{-0.20}$	&	$1.31^{+0.11}_{-0.11}$	\\
& $\tau$ & $$ &								$0.5f$					&	$0.5f$					&	$0.5f$				&											$0.5f$						&	$0.5f$				&	$0.5f$				\\
& l$_{\rm{B}}$ & $$ &						$100f$					&	$100f$					&	$100f$				&											$100f$						&	$100f$				&	$100f$				\\
& l$_{\rm{s}}$ & $$ &						$13.9^{+3.83}_{-3.26}$			&	$21.43^{+3.5}_{-2.57}$	&	$20.92^{+2.62}_{-1.95}$	&											$20.53^{+2.61}_{-2.05}$			&	$20.92^{+2.62}_{-1.95}$	&	$17.99^{+2.18}_{-1.91}$	\\
& kT$_{\rm{bb}}$ & $$ &						$1.05^{+0.05}_{-0.07}$			&	$0.99^{+0.05}_{-0.05}$	&	$0.82^{+0.05}_{-0.07}$	&											$0.94^{+0.03}_{-0.04}$			&	$0.81^{+0.05}_{-0.07}$	&	$1.04^{+0.03}_{-0.04}$	\\
\texttt{bremss} & norm & &						$2.94^{+0.61}_{-0.68}$			&	$5.03^{+0.88}_{-1.27}$	&	$1.39^{+0.08}_{-0.09}$	&											$1.12^{+0.1}_{-0.09}$			&	$1.39^{+0.08}_{-0.09}$	&	$1.56^{+0.25}_{-0.21}$	\\
& kT$_{\rm{bb}}$ & keV &						$4.21^{+0.2}_{-0.14}$			&	$4.15^{+0.28}_{-0.12}$	&	$9.47^{+0.65}_{-0.55}$	&											$9.09^{+0.73}_{-0.64}$			&	$9.47^{+0.65}_{-0.55}$	&	$7.34^{+0.69}_{-0.59}$	\\
\texttt{Confidence} & $\chi^{2}_{\rm{red}}$/d.o.f. & &	0.81/57						&	0.84/63				&	0.97/65				&											1.06/65						&	0.63/68				&	0.86/63				\\
\texttt{Fluxes} & F$_{\rm{bol, abs}}^{a}$ & $10^{-9}$ erg cm$^{-2}$ s$^{-1}$ & 7.4 & 13.1 & 10.8 & 8.9 & 14.6 & 11.4 \\	
& F$_{\rm{BELM}}$/F$_{\rm{bol}}$ & & 0.56 & 0.60 & 0.58 & 0.60 & 0.59 & 0.49 \\
& F$_{\rm{Bremss}}$/F$_{\rm{bol}}$ & & 0.39 & 0.37 & 0.39 & 0.36 & 0.37 & 0.48 \\
& F$_{\rm{Line}}$/F$_{\rm{bol}}$ & & 0.04 & 0.04 & 0.03 & 0.04 & 0.03 & 0.04 \\
\hline
Component & Parameter & Unit & N$^{\underline{o}}$ 7 & N$^{\underline{o}}$ 8 & N$^{\underline{o}}$ 9 & N$^{\underline{o}}$ 10 & N$^{\underline{o}}$ 11 & N$^{\underline{o}}$ 12 \\
\hline
\texttt{phabs} & nH$_{0}$ & $10^{22}$ & 			$3f$ 							&	$3f$					&	$3f$					&											$3f$							&	$3f$					&	$3f$					\\
\texttt{pcfabs} & nH$_{1}$ & $10^{22}$ & 			$12f$						&	$12f$				&	$12f$				&											$12f$						&	$12f$				&	$12f$				\\	
& cov & $$ &								$0.90^{+0.05}_{-0.05}$			&	$0.71^{+0.05}_{-0.05}$	&	$0.92^{+0.03}_{-0.03}$	&											$0.9^{+0.02}_{-0.02}$			&	$0.92^{+0.03}_{-0.06}$	&	$0.81^{+0.06}_{-0.03}$	\\
\texttt{edge(1)} & E$_{\rm{Fe}}^{\rm{ion}}$ & keV &	$9.28f$						&	$9.28f$				&	$9.28f$				&											$9.28f$						&	$9.28f$				&	$9.28f$				\\	
& $\tau(1)_{\rm{Fe}}^{\rm{ion}}$ & $$ & 			$0.17^{+0.04}_{-0.03}$			&	$0.14^{+0.04}_{-0.04}$	&	$0.13^{+0.04}_{-0.05}$	&											$0.21^{+0.02}_{-0.02}$			&	$0.24^{+0.02}_{-0.03}$	&	$0.23^{+0.02}_{-0.02}$	\\
\texttt{edge(2)} & E$_{\rm{Fe}}^{\rm{ion}}$ & keV &	$8.83f$						&	$8.83f$				&	$8.83f$				&											$8.83f$						&	$8.83f$				&	$8.83f$				\\
& $\tau(2)_{\rm{Fe}}^{\rm{ion}}$ & $$ & 			$0.11^{+0.03}_{-0.03}$			&	$0.13^{+0.04}_{-0.04}$	&	$0.13^{+0.05}_{-0.05}$	&											$0.11^{+0.03}_{-0.04}$			&	$0.12^{+0.03}_{-0.03}$	&	$0.10^{+0.03}_{-0.05}$	\\
\texttt{gauss[Fe]} & EW$_{\rm{Fe}}$ & eV			& $370^{+22}_{-21}$		&	$302^{+22}_{-20}$			&	$422^{+47}_{-39}$		&											$310^{+47}_{-37}$				&	$269^{+45}_{-43}$		&	$389^{+50}_{-37}$		\\
& E$_{\rm{Fe}}$ & keV &						$6.42^{+0.03}_{-0.03}$			&	$6.48^{+0.02}_{-0.02}$	&	$6.41^{+0.03}_{-0.03}$	&											$6.42^{+0.03}_{-0.04}$			&	$6.43^{+0.04}_{-0.04}$	&	$6.47^{+0.03}_{-0.03}$	\\
& $\sigma_{\rm{Fe}}$ & keV & 					$0.30^{+0.06}_{-0.06}$			&	$0.14^{+0.07}_{-0.14}$	&	$0.4^{+0.06}_{-0.06}$	&											$0.34^{+0.08}_{-0.08}$			&	$0.34^{+0.09}_{-0.11}$	&	$0.29^{+0.08}_{-0.08}$	\\
\texttt{BELM} & norm & $$ &					$1.26^{+0.19}_{-0.18}$			&	$1.07^{+0.05}_{-0.04}$	&	$1.13^{+0.11}_{-0.08}$	&											$1.09^{+0.05}_{-0.07}$			&	$0.98^{+0.21}_{-0.01}$	&	$0.87^{+0.07}_{-0.06}$	\\
& $\tau$ & $$ &								$0.5f$						&	$0.5f$				&	$0.5f$				&											$0.5f$						&	$0.5f$				&	$0.5f$				\\
& l$_{\rm{B}}$ & $$ &						$38.59f$						&	$100f$				&	$100f$				&											$100f$						&	$100f$				&	$100f$				\\
& l$_{\rm{s}}$ & $$ &						$25.69^{+5.62}_{-3.85}$			&	$22^{+4.63}_{-5.06}$	&	$12.64^{+3.56}_{-2.64}$	&											$10.82^{+2.93}_{-0.82}$			&	$15.18^{+1.8}_{-4.25}$	&	$11.94^{+4.01}_{-1.94}$	\\
& kT$_{\rm{bb}}$ & $$ &						$0.77^{+0.06}_{-0.09}$			&	$1.08^{+0.03}_{-0.05}$	&	$0.95^{+0.07}_{-0.12}$	&											$1.14^{+0.09}_{-0.09}$			&	$>1.13$				&	$>1.15$				\\
\texttt{bremss} & norm & &						$1.78^{+0.18}_{-0.11}$			&	$3.19^{+1.19}_{-1.11}$	&	$4.43^{+0.76}_{-0.74}$	&											$7.24^{+0.43}_{-0.49}$			&	$12.2^{+1.94}_{-4.69}$	&	$5.69^{+18.66}_{-0.64}$	\\
& kT$_{\rm{bb}}$ & keV &						$9.71^{+0.39}_{-0.38}$			&	$4.41^{+0.49}_{-0.25}$	&	$3.28^{+0.08}_{-0.09}$	&											$2.91^{+0.08}_{-0.14}$			&	$2.1^{+0.58}_{-0.14}$	&	$2.61^{+0.24}_{-0.49}$	\\
\texttt{Confidence} & $\chi^{2}_{\rm{red}}$/d.o.f. & &	1.37/67						&	0.92/58				&	0.82/59				&											1.10/61						&	0.89/60				&	0.78/53				\\
\texttt{Fluxes} & F$_{\rm{bol, abs}}^{a}$ & $10^{-9}$ erg cm$^{-2}$ s$^{-1}$ & 10.7 & 10.0 & 7.4 & 8.1 & 8.1 & 6.4 \\	
& F$_{\rm{BELM}}$/F$_{\rm{bol}}$ & & 0.62 & 0.61 & 0.58 & 0.53 & 0.63 & 0.60 \\
& F$_{\rm{Bremss}}$/F$_{\rm{bol}}$ & & 0.34 & 0.36 & 0.37 & 0.43 & 0.34 & 0.35 \\
& F$_{\rm{Line}}$/F$_{\rm{bol}}$ & & 0.04 & 0.04 & 0.05 & 0.04 & 0.04 & 0.05 \\
\hline
\end{tabular}
\end{center}
\begin{list}{}{}
\item[$^{\mathrm{f}}$] Frozen in the fits.
\item[$^{\mathrm{a}}$] Absorbed, bolometric flux (3.5--300 keV) of the model normalized to the PCA data
\end{list}
\end{table*}

\begin{table*}
\contcaption{} \label{wholemodel1}
\begin{center}
\begin{tabular}{lcccccccc}
\hline\hline
Component & Parameter & Unit & N$^{\underline{o}}$ 13 & N$^{\underline{o}}$ 14 & N$^{\underline{o}}$ 15 & N$^{\underline{o}}$ 16 & N$^{\underline{o}}$ 17 & N$^{\underline{o}}$ 18 \\
\hline
\texttt{phabs} & nH$_{0}$ & $10^{22}$ & 			$3f$ 							&	$3f$					&	$3f$					&											$3f$							&	$3f$					&	$3f$					\\
\texttt{pcfabs} & nH$_{1}$ & $10^{22}$ & 			$12f$						&	$9f$					&	$9f$					&											$9f$							&	$9f$					&	$9f$					\\
& cov & $$ &								$0.92^{+0.03}_{-0.04}$			&	$>0.90$				&	$0.90^{+0.05}_{-0.07}$	&											$0.47^{+0.07}_{-0.10}$			&	$0.95^{+0}_{-0.03}$		&	$0.95^{+0}_{-0.01}$		\\
\texttt{edge(1)} & E$_{\rm{Fe}}^{\rm{ion}}$ & keV &	$9.28f$						&	$9.28f$				&	$9.28f$				&											$9.28f$						&	$9.28f$				&	$9.28f$				\\
& $\tau(1)_{\rm{Fe}}^{\rm{ion}}$ & $$ & 			$0.23^{+0.02}_{-0.02}$			&	$0.09^{+0.04}_{-0.04}$	&	$0.14^{+0.04}_{-0.05}$	&											$0.00^{+0.01}_{-0.00}$			&	$0.29^{+0.02}_{-0.02}$	&	$0.18^{+0.03}_{-0.03}$	\\
\texttt{edge(2)} & E$_{\rm{Fe}}^{\rm{ion}}$ & keV &	$8.83f$						&	$8.83f$				&	$8.83f$				&											$8.83f$						&	$8.83f$				&	$8.83f$				\\
& $\tau(2)_{\rm{Fe}}^{\rm{ion}}$ & $$ & 			$0.10^{+0.03}_{-0.03}$			&	$0.17^{+0.05}_{-0.05}$	&	$0.09^{+0.05}_{-0.05}$	&											$0.25^{+0.02}_{-0.02}$			&	$0^{+0.01}_{-0}$		&	$0.04^{+0.04}_{-0.03}$	\\
\texttt{gauss[Fe]} & EW$_{\rm{Fe}}$ & eV 	&		$341^{+34}_{-30}$			&	$288^{+33}_{-18}$			&	$305^{+36}_{-30}$		&											$546^{+23}_{-24}$				&	$416^{+29}_{-26}$		&	$525^{+97}_{-105}$		\\
& E$_{\rm{Fe}}$ & keV &						$6.45^{+0.03}_{-0.03}$			&	$6.42^{+0.03}_{-0.03}$	&	$6.55^{+0.03}_{-0.03}$	&											$6.26^{+0.05}_{-0.02}$			&	$6.63^{+0.03}_{-0.03}$	&	$6.31^{+0.07}_{-0.06}$	\\
& $\sigma_{\rm{Fe}}$ & keV & 					$0.24^{+0.07}_{-0.08}$			&	$0.21^{+0.09}_{-0.12}$	&	$0.26^{+0.08}_{-0.1}$	&											$0.13^{+0.06}_{-0.13}$			&	$0.40^{+0.05}_{-0.05}$	&	$0.74^{+0.09}_{-0.12}$	\\
\texttt{BELM} & norm & $$ &					$1.33^{+0.06}_{-0.05}$			&	$0.73^{+0.04}_{-0.04}$	&	$0.90^{+0.02}_{-0.03}$	&											$0.24^{+0.01}_{-0.01}$			&	$0.65^{+0.03}_{-0.01}$	&	$0.92^{+0.03}_{-0.03}$	\\
& $\tau$ & $$ &								$0.5f$						&	$0.5f$				&	$0.5f$				&											$0.5f$						&	$0.5f$				&	$0.5f$				\\
& l$_{\rm{B}}$ & $$ &						$12.63^{+8.92}_{-5.17}$			&	$29.07^{+42.72}_{-9.96}$	&	$49.15^{+49.34}_{-16.79}$&											$76.4^{+123.6}_{-64.88}$			&	$21.18^{+22.7}_{-10.1}$	&	$20.01^{+11.73}_{-8.49}$	\\
& l$_{\rm{s}}$ & $$ &						$<10.55$						&	$18.18^{+6.65}_{-5.28}$	&	$22.71^{+5.37}_{-5.6}$	&											$10^{+4.62}_{-0}$				&	$10^{+40}_{-0}$		&	$10^{+1.95}_{-0}$		\\
& kT$_{\rm{bb}}$ & $$ &						$1.30^{+0.07}_{-0.11}$			&	$1.03^{+0.05}_{-0.04}$	&	$1.06^{+0.03}_{-0.05}$	&											$1.16^{+0.06}_{-0.05}$			&	$>1.34$				&	$1.18^{+0.08}_{-0.07}$	\\
\texttt{bremss} & norm & &						$7.58^{+1.79}_{-1.9}$			&	$3.62^{+0.89}_{-1.69}$	&	$1.66^{+1.31}_{-1.06}$	&											$0.79^{+0.05}_{-0.25}$			&	$6.57^{+0.17}_{-0.65}$	&	$5.54^{+0.42}_{-0.42}$	\\
& kT$_{\rm{bb}}$ & keV &						$2.13^{+0.25}_{-0.15}$			&	$2.93^{+0.20}_{-0.13}$	&	$4.01^{+1.38}_{-0.38}$	&											$3.85^{+0.32}_{-0.15}$			&	$2.39^{+0.10}_{-0.04}$	&	$2.38^{+0.12}_{-0.15}$	\\
\texttt{Confidence} & $\chi^{2}_{\rm{red}}$/d.o.f. & &	0.97/	72						&	0.96/58				&	1.07/	64				&											1.22/49						&	1.24/	64				&	1.49/	69				\\
\texttt{Fluxes} & F$_{\rm{bol, abs}}^{a}$ & $10^{-9}$ erg cm$^{-2}$ s$^{-1}$ & 7.9 & 5.8 & 7.1 & 2.0 & 5.5 & 6.2 \\	
& F$_{\rm{BELM}}$/F$_{\rm{bol}}$ & & 0.74 & 0.64 & 0.74 & 0.53 & 0.55 & 0.65 \\
& F$_{\rm{Bremss}}$/F$_{\rm{bol}}$ & & 0.22 & 0.33 & 0.23 & 0.41 & 0.40 & 0.30 \\
& F$_{\rm{Line}}$/F$_{\rm{bol}}$ & & 0.03 & 0.03 & 0.04 & 0.06 & 0.04 & 0.06 \\
\hline
Component & Parameter & Unit & N$^{\underline{o}}$ 19 & N$^{\underline{o}}$ 20 & N$^{\underline{o}}$ 21 & N$^{\underline{o}}$ 22 & N$^{\underline{o}}$ 23 & N$^{\underline{o}}$ 24 \\
\hline
\texttt{phabs} & nH$_{0}$ & $10^{22}$ & 			$3f$ 							&	$3f$					&	$3f$					&											$3f$							&	$3f$					&	$3f$					\\
\texttt{pcfabs} & nH$_{1}$ & $10^{22}$ & 			$9f$							&	$9f$					&	$9f$					&											$9f$							&	$9f$					&	$9f$					\\
& cov & $$ &								$0.95^{+0}_{-0.01}$				&	$0.93^{+0.02}_{-0.04}$	&	$0.95^{+0}_{-0.01}$		&											$0.80^{+0.05}_{-0.05}$			&	$0.73^{+0.05}_{-0.06}$	&	$0.95^{+0}_{-0.02}$		\\
\texttt{edge(1)} & E$_{\rm{Fe}}^{\rm{ion}}$ & keV &	$9.28f$						&	$9.28f$				&	$9.28f$				&											$9.28f$						&	$9.28f$				&	$9.28f$				\\
& $\tau(1)_{\rm{Fe}}^{\rm{ion}}$ & $$ & 			$0.15^{+0.02}_{-0.01}$			&	$0.20^{+0.04}_{-0.04}$	&	$0.11^{+0.04}_{-0.04}$	&											$0.02^{+0.04}_{-0.02}$			&	$0.14^{+0.04}_{-0.04}$	&	$0.08^{+0.04}_{-0.04}$	\\
\texttt{edge(2)} & E$_{\rm{Fe}}^{\rm{ion}}$ & keV &	$8.83f$						&	$8.83f$				&	$8.83f$				&											$8.83f$						&	$8.83f$				&	$8.83f$				\\
& $\tau(2)_{\rm{Fe}}^{\rm{ion}}$ & $$ & 			$0.15^{+0.02}_{-0.02}$			&	$0.08^{+0.05}_{-0.05}$	&	$0.14^{+0.05}_{-0.05}$	&											$0.22^{+0.03}_{-0.04}$			&	$0.15^{+0.04}_{-0.04}$	&	$0.18^{+0.04}_{-0.04}$	\\	
\texttt{gauss[Fe]} & EW$_{\rm{Fe}}$ & eV &		$243^{+23}_{-20}$			&	$430^{+33}_{-29}$			&	$364^{+42}_{-35}$		&											$343^{+28}_{-25}$				&	$486^{+15}_{-15}$		&	$323^{+39}_{-32}$		\\
& E$_{\rm{Fe}}$ & keV &						$6.23^{+0.03}_{-0.04}$			&	$6.59^{+0.02}_{-0.02}$	&	$6.44^{+0.03}_{-0.03}$	&											$6.28^{+0.03}_{-0.03}$			&	$6.49^{+0.02}_{-0.01}$	&	$6.40^{+0.03}_{-0.03}$	\\
& $\sigma_{\rm{Fe}}$ & keV & 					$0.25^{+0.07}_{-0.09}$			&	$0.22^{+0.06}_{-0.08}$	&	$0.30^{+0.07}_{-0.08}$	&											$0.20^{+0.07}_{-0.09}$			&	$0.02^{+0.08}_{-0.02}$	&	$0.33^{+0.07}_{-0.07}$	\\
\texttt{BELM} & norm & $$ &					$1.02^{+0.03}_{-0.04}$			&	$0.57^{+0.05}_{-0.05}$	&	$0.60^{+0.03}_{-0.04}$	&											$0.80^{+0.05}_{-0.05}$			&	$0.32^{+0.02}_{-0.02}$	&	$1.31^{+0.05}_{-0.06}$	\\
& $\tau$ & $$ &								$0.5f$						&	$0.5f$				&	$0.5f$				&											$0.5f$						&	$0.5f$				&	$0.5f$				\\
& l$_{\rm{B}}$ & $$ &						$23.4^{+14.86}_{-8.26}$			&	$100f$				&	$100f$				&											$100f$						&	$100f$				&	$100f$				\\
& l$_{\rm{s}}$ & $$ &						$10^{+1.8}_{-0}$				&	$16.21^{+3.19}_{-3.04}$	&	$20.59^{+2.42}_{-2.24}$	&											$22.53^{+2.8}_{-2.63}$			&	$10.47^{+3.14}_{-0.47}$	&	$21.74^{+2.59}_{-2.7}$	\\
& kT$_{\rm{bb}}$ & $$ &						$1.19^{+0.12}_{-0.07}$			&	$1.08^{+0.04}_{-0.05}$	&	$1.01^{+0.03}_{-0.03}$	&											$1.01^{+0.04}_{-0.04}$			&	$1.17^{+0.06}_{-0.06}$	&	$0.99^{+0.03}_{-0.03}$	\\
\texttt{bremss} & norm & &						$10.99^{+0.93}_{-0.47}$			&	$0.86^{+0.20}_{-0.17}$	&	$0.75^{+0.15}_{-0.12}$	&											$1.64^{+0.33}_{-0.3}$			&	$0.87^{+0.06}_{-0.19}$	&	$1.88^{+0.51}_{-0.42}$	\\
& kT$_{\rm{bb}}$ & keV &						$2.41^{+0.06}_{-0.11}$			&	$6.45^{+0.76}_{-0.49}$	&	$6.39^{+0.63}_{-0.46}$	&											$5.45^{+0.47}_{-0.34}$			&	$5.26^{+0.25}_{-0.14}$	&	$5.51^{+0.60}_{-0.40}$	\\
\texttt{Confidence} & $\chi^{2}_{\rm{red}}$/d.o.f. & &	1.29/75						&	1.23/57				&	1.13/60				&											0.84/61						&	0.91/56				&	1.01/62				\\
\texttt{Fluxes} & F$_{\rm{bol, abs}}^{a}$ & $10^{-9}$ erg cm$^{-2}$ s$^{-1}$ & 8.4 & 4.6 & 4.7 & 7.3 & 2.8 & 10.1 \\	
& F$_{\rm{BELM}}$/F$_{\rm{bol}}$ & & 0.52 & 0.60 & 0.66 & 0.60 & 0.48 & 0.67 \\
& F$_{\rm{Bremss}}$/F$_{\rm{bol}}$ & & 0.45 & 0.36 & 0.30 & 0.35 & 0.47 & 0.29 \\
& F$_{\rm{Line}}$/F$_{\rm{bol}}$ & & 0.03 & 0.04 & 0.04 & 0.04 & 0.05 & 0.04 \\
\hline
\end{tabular}
\end{center}
\begin{list}{}{}
\item[$^{\mathrm{f}}$] Frozen in the fits.
\item[$^{\mathrm{a}}$] Absorbed, bolometric flux (3.5--300 keV) of the model normalized to the PCA data
\end{list}
\end{table*}

\begin{table*}
\contcaption{} \label{wholemodel1}
\begin{center}
\begin{tabular}{lcccccccc}
\hline\hline
Component & Parameter & Unit & N$^{\underline{o}}$ 25 & N$^{\underline{o}}$ 26 & N$^{\underline{o}}$ 27 \\
\hline
\texttt{phabs} & nH$_{0}$ & $10^{22}$ & 			$3f$ 							&	$3f$					&	$3f$					\\
\texttt{pcfabs} & nH$_{1}$ & $10^{22}$ & 			$9f$							&	$9f$					&	$9f$					\\	
& cov & $$ &								$0.85^{+0.05}_{-0.05}$			&	$0.70^{+0.06}_{-0.06}$	&	$0.81^{+0.05}_{-0.05}$	\\
\texttt{edge(1)} & E$_{\rm{Fe}}^{\rm{ion}}$ & keV &	$9.28f$						&	$9.28f$				&	$9.28f$				\\	
& $\tau(1)_{\rm{Fe}}^{\rm{ion}}$ & $$ & 			$0.10^{+0.04}_{-0.04}$			&	$0.16^{+0.05}_{-0.05}$	&	$0.17^{+0.04}_{-0.04}$	\\
\texttt{edge(2)} & E$_{\rm{Fe}}^{\rm{ion}}$ & keV &	$8.83f$						&	$8.83f$				&	$8.83f$				\\
& $\tau(2)_{\rm{Fe}}^{\rm{ion}}$ & $$ & 			$0.17^{+0.04}_{-0.04}$			&	$0.13^{+0.05}_{-0.05}$	&	$0.08^{+0.04}_{-0.04}$	\\
\texttt{gauss[Fe]} & EW$_{Fe}$ & eV  &			$316^{+27}_{-24}$				&	$479^{+27}_{-26}$		&	$299^{+33}_{-29}$		\\
& E$_{\rm{Fe}}$ & keV &						$6.47^{+0.02}_{-0.03}$			&	$6.51^{+0.02}_{-0.02}$	&	$6.47^{+0.03}_{-0.03}$	\\
& $\sigma_{\rm{Fe}}$ & keV & 					$0.20^{+0.06}_{-0.08}$			&	$0.13^{+0.06}_{-0.13}$	&	$0.31^{+0.07}_{-0.07}$	\\
\texttt{BELM} & norm & $$ &					$1.75^{+0.12}_{-0.11}$			&	$0.64^{+0.16}_{-0.19}$	&	$1.77^{+0.16}_{-0.16}$	\\
& $\tau$ & $$ &								$0.5f$						&	$0.5f$				&	$0.5f$				\\
& l$_{\rm{B}}$ & $$ &						$38.59f$						&	$100f$				&	$100f$				\\
& l$_{\rm{s}}$ & $$ &						$18.07^{+2.2}_{-2.11}$			&	$18^{+7.92}_{-4.41}$	&	$17.82^{+2.31}_{-2.03}$	\\
& kT$_{\rm{bb}}$ & $$ &						$1.04^{+0.04}_{-0.05}$			&	$1.12^{+0.05}_{-0.05}$	&	$0.99^{+0.04}_{-0.05}$	\\
\texttt{bremss} & norm & &						$3.01^{+0.53}_{-0.48}$			&	$0.90^{+0.16}_{-0.11}$	&	$1.99^{+0.34}_{-0.27}$	\\
& kT$_{\rm{bb}}$ & keV &						$6.27^{+0.46}_{-0.39}$			&	$9.65^{+1.53}_{-1.2}$	&	$7.31^{+0.72}_{-0.63}$	\\
\texttt{Confidence} & $\chi^{2}_{\rm{red}}$/d.o.f. & &	1.27/64						&	1.02/53				&	0.89/64				\\
\texttt{Fluxes} & F$_{\rm{bol, abs}}^{a}$ & $10^{-9}$ erg cm$^{-2}$ s$^{-1}$ & 15.0 & 6.7 & 13.6 \\	
& F$_{\rm{BELM}}$/F$_{\rm{bol}}$ & & 0.58 & 0.52 & 0.63 \\
& F$_{\rm{Bremss}}$/F$_{\rm{bol}}$ & & 0.38 & 0.44 & 0.34 \\
& F$_{\rm{Line}}$/F$_{\rm{bol}}$ & & 0.03 & 0.05 & 0.03 \\
\hline
\end{tabular}
\end{center}
\begin{list}{}{}
\item[$^{\mathrm{f}}$] Frozen in the fits.
\item[$^{\mathrm{a}}$] Absorbed, bolometric flux (3.5--300 keV) of the model normalized to the PCA data
\end{list}
\end{table*}

\begin{table*}
\caption{\texttt{phabs} $\times$ \texttt{pcfabs} $\times$ \texttt{edge(1)} $\times$ \texttt{edge(2)}  $\times$ (\texttt{linemodel} + \texttt{belm} + \texttt{bremss})} \label{wholemodel2}
\begin{center}
\begin{tabular}{lcccccccc}
\hline\hline
Component & Parameter & Unit & N$^{\underline{o}}$ 1 & N$^{\underline{o}}$ 2 & N$^{\underline{o}}$ 3 \\
\hline
\texttt{phabs} & nH$_{0}$ & $10^{22}$ & 			$3.22^{+0.12}_{-0.11}$ 			&	$4.18^{+0.09}_{-0.09}$	&	$3.00^{+0.10}_{-0.10}$	\\
\texttt{pcfabs} & nH$_{1}$ & $10^{22}$ & 			$10.44^{+0.52}_{-0.49}$			&	$16.2^{+0.8}_{-0.8}$		&	$8.98^{+0.34}_{-0.32}$	\\	
& cov & $$ &								$0.79^{+0.01}_{-0.01}$			&	$0.73^{+0.01}_{-0.01}$	&	$0.85^{+0.01}_{-0.01}$	\\
\texttt{edge(1)} & E$_{\rm{Fe}}^{\rm{ion}}$ & keV &	$9.28f$						&	$9.28f$				&	$9.28f$				\\	
& $\tau(1)_{\rm{Fe}}^{\rm{ion}}$ & $$ & 			$0.15^{+0.02}_{-0.03}$			&	$0.14^{+0.03}_{-0.03}$	&	$0.09^{+0.03}_{-0.03}$	\\
\texttt{edge(2)} & E$_{\rm{Fe}}^{\rm{ion}}$ & keV &	$8.83f$						&	$8.83f$				&	$8.83f$				\\
& $\tau(2)_{\rm{Fe}}^{\rm{ion}}$ & $$ & 			$0.11^{+0.03}_{-0.02}$			&	$0.10^{+0.02}_{-0.02}$	&	$0.17^{+0.03}_{-0.03}$	\\
\texttt{BELM} & norm & $$ &					$1.02^{+0.03}_{-0.03}$			&	$1.39^{+0.04}_{-0.04}$	&	$0.77^{+0.01}_{-0.01}$	\\
& $\tau$ & $$ &								$0.5f$						&	$0.5f$				&	$0.5f$				\\
& l$_{\rm{B}}$ & $$ &						$>122$						&	$>185$				&	$101^{+43}_{-32}$		\\
& l$_{\rm{s}}$ & $$ &						$18.9^{+1.4}_{-0.8}$				&	$23.6^{+0.8}_{-0.8}$		&	$<11.0$				\\
& kT$_{\rm{bb}}$ & $$ &						$1.00^{+0.02}_{-0.02}$			&	$0.98^{+0.01}_{-0.01}$	&	$1.17^{+0.04}_{-0.03}$	\\
\texttt{bremss} & norm & &						$1.22^{+0.09}_{-0.09}$			&	$1.02^{+0.09}_{-0.08}$	&	$4.48^{+0.14}_{-0.16}$	\\
& kT$_{\rm{bb}}$ & keV &						$6.81^{+0.27}_{-0.2}$			&	$7.65^{+0.33}_{-0.31}$	&	$2.89^{+0.04}_{-0.05}$	\\
\texttt{Confidence} & $\chi^{2}_{\rm{red}}$/d.o.f. & &	1.48/225						&	1.57/	291				&	1.61/217				\\
\texttt{Fluxes} & F$_{\rm{bol, abs}}^{a}$ & $10^{-9}$ erg cm$^{-2}$ s$^{-1}$ & 10.9 & 10.5 & 5.7 \\	
& F$_{\rm{BELM}}$/F$_{\rm{bol}}$ & & 0.62 & 0.72 & 0.55 \\
& F$_{\rm{Bremss}}$/F$_{\rm{bol}}$ & & 0.34 & 0.25 & 0.41 \\
& F$_{\rm{Line}}$/F$_{\rm{bol}}$ & & 0.04 & 0.03 & 0.04 \\
\hline
\end{tabular}
\end{center}
\begin{list}{}{}
\item[$^{\mathrm{f}}$] Frozen in the fits.
\item[$^{\mathrm{a}}$] Absorbed, bolometric flux (3.5--300 keV) of the model normalized to the PCA data
\end{list}
\end{table*}

\label{lastpage}

\end{document}